\documentclass[%
reprint,
superscriptaddress,
nofootinbib,
amsmath,amssymb,
aps,
prb,
]{revtex4-2}

\usepackage{physics}
\usepackage{xcolor}
\usepackage{graphicx}
\usepackage{bm}
\usepackage{hyperref}

\hypersetup{
colorlinks=true,
linkcolor=blue,
urlcolor=cyan,
citecolor=red,
}
\usepackage[normalem]{ulem}
\usepackage{verbatim}
\usepackage{enumitem}

\newcommand{\<}{\langle}

\renewcommand{\>}{\rangle}
\renewcommand{\(}{\left(}
\renewcommand{\)}{\right)}
\renewcommand{\[}{\left[}
\renewcommand{\]}{\right]}
\renewcommand{\v}[1]{\boldsymbol{#1}} 

\newcommand{\bs}[1]{\boldsymbol{#1}}

\begin{document}

\title{Efficient sampling of noisy shallow circuits via monitored unraveling} 

\author{Zihan Cheng}
\affiliation{Department of Physics, University of Texas at Austin, Austin, TX 78712, USA}

\author{Matteo Ippoliti}
\affiliation{Department of Physics, Stanford University, Stanford, CA 94305, USA}
\affiliation{Department of Physics, University of Texas at Austin, Austin, TX 78712, USA}

\date{\today}

\begin{abstract}
We introduce a classical algorithm for sampling the output of shallow, noisy random circuits on two-dimensional qubit arrays. The algorithm builds on the recently-proposed ``space-evolving block decimation'' (SEBD)~[\href{https://journals.aps.org/prx/abstract/10.1103/PhysRevX.12.021021}{Napp {\it et al}, PRX {\bf 12}, 021021 (2022)}] and extends it to the case of noisy circuits.
SEBD is based on a mapping of 2D unitary circuits to 1D {\it monitored} ones, which feature measurements alongside unitary gates; it exploits the presence of a measurement-induced entanglement phase transition to achieve efficient (approximate) sampling below a finite critical depth $T_c$. 
Our noisy-SEBD algorithm unravels the action of noise into measurements, further lowering entanglement and enabling efficient classical sampling up to larger circuit depths. 
We analyze a class of physically-relevant noise models (unital qubit channels) within a two-replica statistical mechanics treatment, finding weak measurements to be the optimal (i.e. most disentangling) unraveling. 
We then locate the noisy-SEBD complexity transition as a function of circuit depth and noise strength in realistic circuit models.
As an illustrative example, we show that circuits on heavy-hexagon qubit arrays with noise rates of $\approx 2\%$ per CNOT, based on IBM Quantum processors, can be efficiently sampled up to a depth of 5 iSWAP (or 10 CNOT) gate layers.
Our results help sharpen the requirements for practical hardness of simulation of noisy hardware.
\end{abstract}

\maketitle

\tableofcontents


\section{Introduction \label{sec:intro} }

Before quantum computers can do anything useful, they must be able to reliably beat classical computers at some task, ideally one that is quantifiable, well-understood theoretically, and has reasonable experimental requirements. 
{\it Random circuit sampling} (RCS) has emerged as one of the leading candidates for this role: it is well-suited to the architectures of present-day gate-based quantum processors, and---at least in the ideal case of noiseless computation---its hardness is firmly established in complexity theory~\cite{aaronson_complexity-theoretic_2017,boixo_characterizing_2018,neill_blueprint_2018,bouland_complexity_2019,movassagh_quantum_2020}. 
As a result, it has become the focus of pioneering experimental efforts in the last few years~\cite{arute_quantum_2019,wu_strong_2021,zhu_quantum_2022,morvan_phase_2023}.
All such experiments, however, are by necessity carried out on present-day noisy, intermediate scale quantum (NISQ) processors, where the question of classical simulation hardness is much more nuanced.  
This has spurred much interest in the exploration of how noise affects the boundary between ``easy'' and ``hard'' simulation problems. 

It is now established that RCS in the presence of a finite noise rate can be simulated in polynomial time~\cite{aharonov_polynomial-time_2023} in the regime of anticoncentration (i.e., informally, at sufficiently large depth~\cite{dalzell_anticoncentrate_2022}). However the algorithm of Ref.~\cite{aharonov_polynomial-time_2023} is not practical, leaving open the question of simulability of finite-sized, noisy RCS experiments with reasonable classical resources. 
This issue is subtle and depends on many variables, such as circuit architecture, size and depth, details of the noise models, choice of target metrics, etc.
A powerful classical approach is based on {\it tensor networks}, which leverage limited entanglement and work best in 1D; for this reason, experiments have focused on 2D qubit arrays and picked highly-entangling gate sets. Large circuit depth also generates more entanglement and thus makes simulation harder.
However, in practice, depth is limited by the presence of noise, as more gates also cause the accumulation of more errors.
Additionally, the presence of noise in the quantum experiment lowers the bar for classical simulation: an apples-to-apples comparison requires that we tolerate similar levels of error from the classical algorithm as well. This opens the door to various strategies based on tensor networks that can outperform sufficiently-noisy quantum computers~\cite{zhou_what_2020,kalachev_classical_2021,ayral_density-matrix_2023}.

\begin{figure*}
	\centering 
	\includegraphics[width = \textwidth]{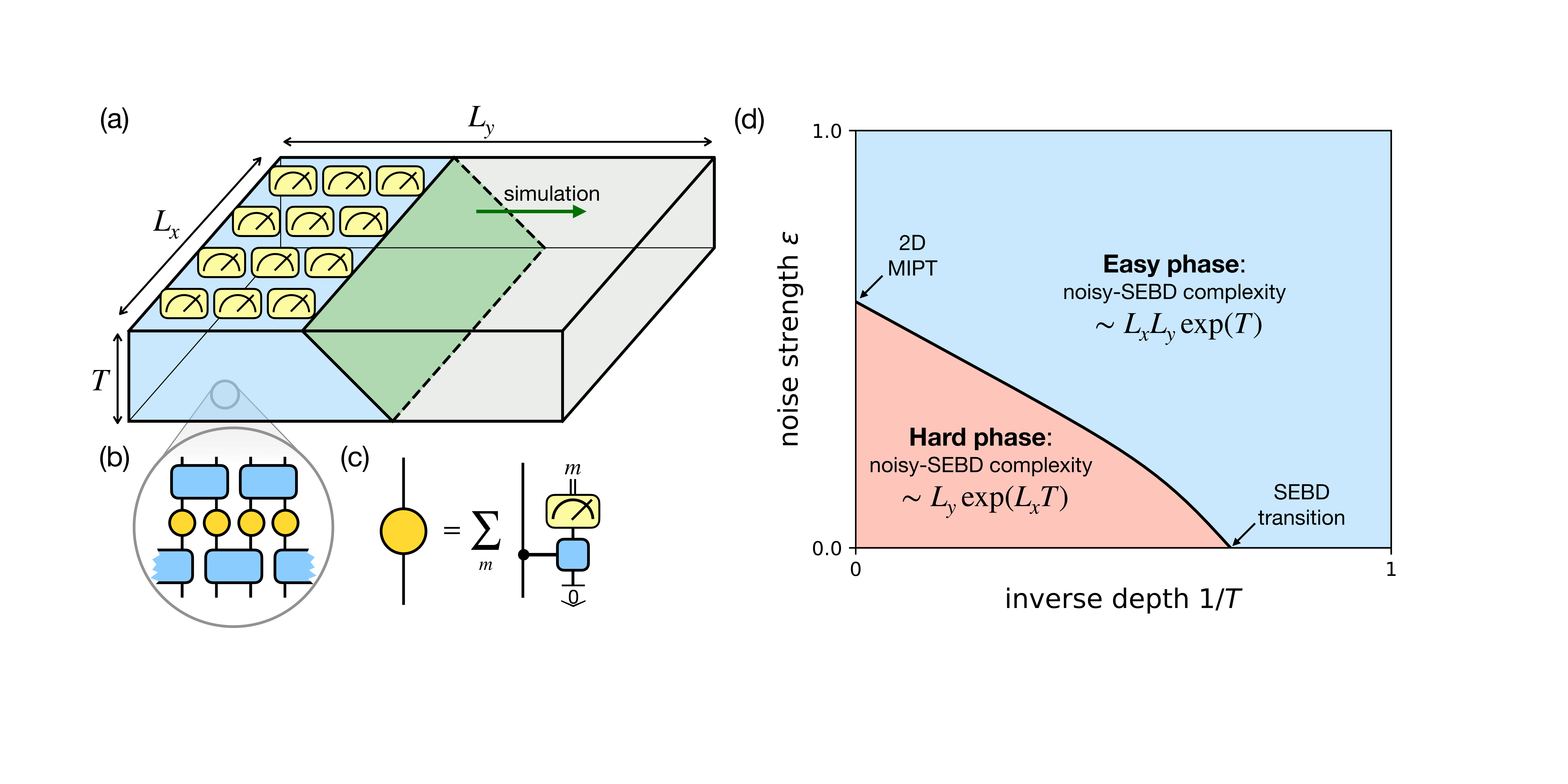}
	\caption{Main ideas of this work. 
	(a) Sampling of 2D shallow unitary circuits. The qubit array has linear dimensions $L_x$, $L_y$ and the circuit has depth $T$. The SEBD algorithm is based on an MPS simulation carried out along a spatial direction (e.g. $L_y$), approximating the wavefunction of a 1D subsystem of qubits on a light cone surface (green). It exploits the measurements on the top boundary of the circuit to disentangle the wavefunction. 
	(b) We consider noisy circuits with uncorrelated local noise: unitary gates (blue rectangles) are interspersed with single-qubit noise channels (orange circles). 
	(c) The noise channels are unraveled as measurements (generically weak), represented here as interactions with a fresh ancilla which is then measured. These fictitious measurements can further disentangle the wavefunction. 
	(d) Qualitative sketch of the complexity phase diagram of the noisy-SEBD algorithm. An entanglement phase transition separates an ``easy'' phase (low depth and/or strong noise), where the computational cost of sampling the noisy circuit output via noisy-SEBD is $\sim L_x L_y \exp(T)$, from a ``hard'' phase where the same cost becomes $\sim L_y \exp(L_x T)$.
	The location of the phase boundary is model-dependent. The line $\varepsilon = 0$ yields the finite-depth complexity transition of Ref.~\cite{napp_efficient_2022} (noiseless SEBD) while the $1/T = 0$ line (which stands for $T = O(L)$) yields the standard measurement-induced phase transition in two spatial dimensions~\cite{turkeshi_measurement-induced_2020}. We conjecture the scaling $\varepsilon_c(T) \sim \varepsilon_{c,2D} + O(1/T)$ at large $T$.
	We emphasize that this is a transition in the complexity of the noisy-SEBD algorithm, {\it not} of the sampling task itself~\cite{aharonov_polynomial-time_2023}.
	\label{fig:idea}
	}
\end{figure*}

Characterizing the practical hardness of noisy RCS problems is thus a pressing question in quantum information science. It is also closely related to recent developments in nonequilibrium many-body physics regarding dynamical phases of quantum information in open systems. 
In particular, the fate of unitary circuits that are sampled {\it during} the dynamics (a special case of open-system evolution) has attracted much interest due to the discovery of entanglement phases that occur as a function of the measurement rate~\cite{li_quantum_2018,li_measurement-driven_2019, skinner_measurement-induced_2019,chan_unitary-projective_2019,szyniszewski_entanglement_2019,cao_entanglement_2019,choi_quantum_2020,bao_theory_2020, jian_measurement-induced_2020,turkeshi_measurement-induced_2020,gullans_dynamical_2020,gullans_scalable_2020,zabalo_critical_2020,bao_symmetry_2021,li_conformal_2021,li_entanglement_2023,alberton_entanglement_2021,lavasani_measurement-induced_2021,ippoliti_entanglement_2021,nahum_measurement_2021,sang_measurement-protected_2021,lavasani_topological_2021,ippoliti_postselection-free_2021,lu_spacetime_2021,ippoliti_fractal_2022,zabalo_operator_2022,weinstein_measurement-induced_2022,noel_measurement-induced_2022,koh_measurement-induced_2023,hoke_measurement-induced_2023}.
The emergence of a phase with limited entanglement (area-law) at high measurement rate in these circuits has been used to develop an efficient algorithm for sampling the output of shallow unitary circuits in 2D~\cite{napp_efficient_2022}. The algorithm, dubbed ``space-evolving block decimation'' (SEBD) and sketched in Fig.~\ref{fig:idea}(a), works at depths $T$ below a finite critical depth $T_c$ (model-dependent, but $\approx 4-5$ in typical models). Circuits with $T = 3$ in 2D are capable of universal quantum computation and thus hard to sample in the worst case~\cite{terhal_adaptive_2004}, making this result surprising. 
Another very recent development in dynamical phases of quantum information is the discovery of a sharp noise-induced phase transition in RCS~\cite{morvan_phase_2023,ware_sharp_2023}, which was identified at a fixed number of errors per circuit layer (i.e. noise strength $\varepsilon \sim 1/N$, $N$ being the number of qubits). In the strong-noise phase, it was argued~\cite{morvan_phase_2023} that the processor's output can be classically ``spoofed'', while in the weak-noise phase simulation is conjectured to be practically hard. 

In this work, we consider the problem of noisy RCS on shallow 2D circuits, Fig.~\ref{fig:idea}(a-b), where the results of Ref.~\cite{aharonov_polynomial-time_2023} are not applicable. Unlike other works, our goal is not to sample from the ideal output within some tolerance determined by the noise; instead, we aim to accurately sample the noisy output itself---a closely related but different problem.
To this end we develop a classical algorithm, dubbed noisy-SEBD, and show that it undergoes a complexity phase transition (from quadratic to exponential in the linear size of the system) as a function of circuit depth $T$ and noise strength $\varepsilon$. 
The physical principle behind the algorithm is that noise can be viewed as a sequence of (fictitious) measurements done by the environment on the system, Fig.~\ref{fig:idea}(c). Measurements can lower the amount of entanglement in the system and drive a phase transition to an area-law entangled phase, where tensor network simulation is efficient; as a consequence, noise can drive a phase transition in the complexity of noisy-SEBD. 
Since the unraveling of noise into measurements is not unique, we are free to optimize it in order to lower this threshold as much as possible. In this work we use a two-replica statistical mechanics model to optimize the unraveling, finding {\it weak measurements} to be more disentangling than stochastic projective measurements; the effect of this optimized choice lowers the threshold noise rate by as much as a factor of $\approx 2$, thus significantly expanding the ``easy'' phase.

Combined with the depth-induced complexity transition in the original (noiseless) SEBD algorithm~\cite{napp_efficient_2022}, this defines a phase boundary in the space of circuit depth $T$ and noise strength $\varepsilon$, sketched in Fig.~\ref{fig:idea}(d). 
This adds to our growing understanding of the boundaries of ``practical'' simulability of noisy quantum systems. 
Moreover, it places sharp constraints on the possibility of achieving beyond-classical computation by scaling RCS experiments in space only---i.e., by growing quantum processor size at fixed circuit depth. This highlights the importance of further improvements in error rates of NISQ hardware. 

The paper is structured as follows.
Sec.~\ref{sec:review} reviews background material on random circuit sampling, matrix-product state simulation methods, the SEBD algorithm and monitored dynamics. 
In Sec.~\ref{sec:unraveling} we discuss the unraveling of noise into monitored trajectories and the choice of entanglement-optimal unravelings, including explicit solutions for unital qubit channels. 
Sec.~\ref{sec:nsebd} presents the noisy-SEBD algorithm, numerical simulations of its complexity phase transition, and an illustrative application to circuits based on IBM Quantum's heavy-heaxagon qubit arrays.
We conclude in Sec.~\ref{sec:discussion} by summarizing our results, their implications and connections with other works, and directions for future research.

\section{Background \label{sec:review}}

\subsection{Random circuit sampling \label{sec:rcs}} 

Random circuit sampling (RCS) has emerged as a leading candidate for early demonstrations of quantum computational supremacy, combining good fit with existing hardware capabilities and robust complexity-theoretic arguments for classical hardness~\cite{aaronson_complexity-theoretic_2017,boixo_characterizing_2018,neill_blueprint_2018,bouland_complexity_2019,movassagh_quantum_2020}. 
The idea is to draw a random instance $U$ from an ensemble of local unitary circuits of depth $T$, run it on a quantum computer prepared in the initial state $\ket{\v{0}} \equiv \ket{0}^{\otimes N}$, and measure the state of each qubit in the computational basis, obtaining a bitstring $\v{z}\in\{0,1\}^N$.
Ideally, this process samples bitstrings from the probability distribution $P_U(\v{z}) = |\bra{\v{z}} U \ket{\v{0}}|^2$, which is computationally hard to do for classical computers. 
Intuitively, this is due to the production of extensive entanglement in the system over the course of typical instances of the unitary evolution $U$~\cite{nahum_quantum_2017} which causes tensor network classical algorithms to fail. At the same time, sufficiently generic ensembles of unitary gates in $U$ ensure that various other strategies for efficient classical simulation (such as stabilizers~\cite{aaronson_improved_2004} or matchgates~\cite{valiant_quantum_2002,terhal_classical_2002,jozsa_matchgates_2008}) are not viable. 

These theoretical insights have motivated pioneering experimental efforts in the past few years to demonstrate RCS-based quantum computational supremacy on present-day NISQ hardware~\cite{arute_quantum_2019,wu_strong_2021,zhu_quantum_2022,morvan_phase_2023}. Verification of successful RCS is nontrivial. The experiments employ a linear cross-entropy diagnostic 
\begin{equation}
	{\sf XEB} = 2^N \sum_{\v z} p_{\rm exp}(\v z) p_U(\v z) - 1, \label{eq:XEB}
\end{equation}
where $p_U(\v z)$ is the previously-defined ideal distribution (to be computed classically), and $p_{\rm exp}(\v z)$ is the distribution of bitstrings obtained from the experiment, which generally differs from the ideal one due to imperfect implementation and uncontrolled noise. This quantity is convenient as it can be estimated by sampling from the experiment: 
\begin{equation}
	{\sf XEB} = 2^N \< p_U(\v z) \>_{\v{z} \sim p_{\rm exp}} - 1. \label{eq:XEB_sample}
\end{equation}
Furthermore, it is designed in such a way that ${\sf XEB} = 1$ if the experiment is perfect ($p_{\rm exp} = p_U$, assuming the Porter-Thomas distribution), while ${\sf XEB} = 0$ if it is completely noisy ($p_{\rm exp}(\v z) = 2^{-N}$ for all $\v z$).
In real-world conditions, with finite noise per gate, it has been argued that ${\sf XEB} \sim f^{NT}$ in the regime of interest to the experiment~\cite{arute_quantum_2019}, where $f$ is the average fidelity per gate (more detailed results on the conversion of local noise into global depolarizing noise have been obtained subsequently~\cite{deshpande_tight_2022,dalzell_random_2021}). 
This lowers the bar for classical algorithms: they do not need to achieve a perfect score ${\sf XEB} = 1$, but only to exceed the fidelity of the NISQ experiment. 

This has led to a flurry of activity in the past few years to develop approximate classical algorithms that can efficiently simulate RCS with ${\sf XEB}$ scores and runtimes comparable to those of the NISQ experiments~\cite{zhou_what_2020,kalachev_classical_2021,gao_limitations_2021,pan_solving_2022,ayral_density-matrix_2023}.
Furthermore, it was recently shown that there exists a polynomial-time algorithm for RCS with constant noise per gate~\cite{aharonov_polynomial-time_2023}, which uses a Feynman path-integral representation for the amplitudes $\bra{\v{z}} U \ket{\v{0}}$ wherein noise damps the contribution of most paths. While this settles the complexity of noisy RCS formally, at least in the regime of anticoncentration~\cite{dalzell_anticoncentrate_2022}, the algorithm comes with a very large exponent and is expected to be impractical at the relevant noise strengths.
Thus the {\it practical} issue of efficient classical simulation of finite-sized noisy circuits remains open. 

\subsection{MPS simulation and the entanglement barrier \label{sec:mps_review}}

The interplay of entanglement and noise and its effects on the complexity of classical simulation become especially transparent in the case of matrix-product state (MPS) simulations of random circuits, limited to 1D (or quasi-1D) geometries~\cite{perez-garcia_matrix_2007,verstraete_matrix_2008,cirac_matrix_2021}.
The idea is to represent a wavefunction $\ket{\psi} = \sum_{\v z} c_{\v z}\ket{\v z}$ as a product of three-index tensors via $c_{\v z} = \Tr(A^{z_1}_1 A^{z_2}_2\cdots A^{z_N}_N)$, where each $A^z_i$ is a $\chi\times \chi$ matrix ($i$ labels position in the chain, $z$ is the physical state $\ket{z}$, and two virtual indices are implicit). 
The {\it bond dimension} $\chi$ is a cutoff on the Schmidt rank of the state $\ket{\psi}$, which makes the state classically representable. 
The MPS ansatz allows approximate simulations with high accuracy whenever the entanglement entropy\footnote{In fact compression into MPS form depends on the the behavior of small Schmidt eigenvalues which is captured by Renyi entropies $S_n$ with $n < 1$. In typical random circuits the von Neumann and Renyi entropies have similar scaling for all values of $n$ away from 0.} of $\ket{\psi}$ about each cut obeys $S \ll \ln(\chi)$.
Given the linear growth of entanglement in random circuits~\cite{nahum_quantum_2017}, the MPS method enables accurate simulation for short circuit depths $T \lesssim \ln(\chi)$, after which the truncation of bond dimension incurs a large error. 
One can nonetheless carry out finite-$\chi$ simulations for larger depths; the effect of MPS truncation error on the fidelity with the true state is found to be qualitatively similar to the effect of noise in the quantum experiments~\cite{zhou_what_2020,ayral_density-matrix_2023}. Thus if noise strength is large enough, classical MPS simulations may beat NISQ experiments at the task of approximating a given ideal random circuit.

A different task is to classically simulate (or sample from) noisy circuits themselves. Here, classical simulation needs to incorporate noise and thus mixed states. This can be accomplished by using matrix-product operators (MPO): a mixed state $\rho = \sum_{\v{z}, \v{z}'} c_{\v{z}, \v{z}'} \ket{\v z}\bra{\v{z}'}$ is represented as a product of four-index tensors via $c_{\v{z}\v{z}'} = \Tr(A_1^{z_1,z_1'} A_2^{z_2,z_2'} \cdots A_N^{z_N, z_N'})$, where each $A_i^{z,z'}$ is a $\chi \times \chi$ matrix. 
(A representation that guarantees positivity is given by {\it matrix-product density operators} (MPDO)~\cite{verstraete_matrix_2004}.)
In the presence of noise and unstructured random gates, the unique steady state is expected to be the maximally mixed state $\rho = \mathbb{I}/2^N = \bigotimes_i (\mathbb{I}/2)_i$, which is manifestly disentangled, and can be written as an MPO of bond dimension $\chi = 1$ (i.e. the tensors $A^{z,z'}_i \equiv \delta_{z,z'}/2$ carry no virtual indices). 
Thus entanglement grows initially due to random unitary interactions, $S \sim T$; this persists until the effects of noise are felt, at depth $T \sim 1/p$, $p$ being noise strength; after that, $S$ decreases to zero. 
Thus to accurately simulate the noisy dynamics one has to overcome an ``entanglement barrier'' of height $S \sim 1/p$, with computational cost $\sim {\rm poly}(N, \exp(1/p))$~\cite{noh_efficient_2020,Cheng_simulating_2021}.
While polynomial in the system size $N$, for realistic noise strength $p \sim 10^{-2}$ the cost can still be prohibitively large.
Moreover, the efficient MPO simulation only applies to 1D; in higher dimension, small entanglement generally does not guarantee efficient simulation~\cite{cirac_matrix_2021}.

\subsection{2D shallow circuits and the SEBD algorithm \label{sec:sebd_review}}

Due to MPS simulations, 1D circuit architectures require large depths (diverging faster than $\log(N)$) for hardness. On the contrary, 2D circuits could be hard already for finite depth $T$, as tensor network methods in two or more dimensions are generally not efficient. 
As hardness of simulation generally scales exponentially in the treewidth of the tensor network, experimental RCS works in two-dimensional circuits~\cite{arute_quantum_2019,wu_strong_2021,zhu_quantum_2022,morvan_phase_2023} set $T \sim \sqrt{N}$ to maximize classical hardness. Still, it is natural to ask at what depth $T$ the classical simulation would become hard (asymptotically in large $N$).

It is straightforward to note that $T \leq 2$ is easy, as the output state $U\ket{\v{0}}$ is given by a tensor product of decoupled dimers ($T=1$) or one-dimensional subsystems each hosting an MPS of finite bond dimension ($T=2$).
However, starting at depth $T = 3$, it is possible to prepare states whose exact simulation is provably hard~\cite{terhal_adaptive_2004}.

Surprisingly, Ref.~\cite{napp_efficient_2022} has shown that {\it approximate sampling} from 2D circuits up to a finite depth $T_c\geq 3$ is in fact possible with polynomial time classical algorithms. 
One of the methods they propose, dubbed {\it space-evolving block decimation} (SEBD), is based on reducing the sampled (2+1)D circuit to a (1+1)D circuit featuring measurements alongside unitary gates. 
Below a critical depth $T_c$, the state that needs to be simulated classically obeys an area-law for entanglement~\cite{eisert_colloquium_2010} and can thus be accurately simulated via MPS methods.
The emergence of this low-entanglement phase is an instance of a general phenomenon taking place in {\it monitored dynamics}, i.e. time evolution that combines unitary interactions and measurements, which we review next.

\subsection{Monitored dynamics\label{sec:mipt_review}}

Measurements can disentangle a quantum state. Given a many-body wavefunction $\ket{\psi}$, measuring a qubit $i$ in the computational basis leaves behind a product state between $i$ and the rest of the system: $\propto \ket{z}_i \otimes \braket{z}{\psi}_{\neg i}$ ($z\in\{0,1\}$ is the measurement outcome, obtained randomly from the Born rule, and $\neg i$ denotes the rest of the system). This not only disentangles $i$, but may also destroy entanglement globally---as an extreme example, measuring any one qubit of a GHZ state $\ket{\psi} = \frac{1}{\sqrt{2}}(\ket{0}^{\otimes N} + \ket{1}^{\otimes N})$ in the computational basis leaves behind a global product state.
At the same time, in systems with local interactions, entanglement is generated locally. This asymmetry between entanglement creation and destruction suggests that monitored dynamics, featuring a finite rate $p$ of measurements alongside local unitary interactions, should generically lead to states with low entanglement, for any rate $p>0$~\cite{chan_unitary-projective_2019}. 

Surprisingly, it was found that monitored dynamics can instead successfully stabilize a highly entangled phase as long as the rate of measurement $p$ is below a critical threshold $p_c > 0$~\cite{skinner_measurement-induced_2019,li_quantum_2018,li_measurement-driven_2019}. 
The phases are characterized by the structure of entanglement in {pure} output states of the dynamics, $\ket{\psi_{\v m}}$, which are indexed by the measurement record $\v{m}$ (the sequence of classical outcomes collected during the dynamics).
In particular, the scaling of entanglement entropy $S_A$ for a subsystem $A$ in these states undergoes a transition from an area-law $S_A \sim |\partial A|$ ($\partial A$ is the boundary of $A$) to a volume-law $S_A \sim |A|$.
These scalings are generally washed out in the mixed state $\rho = \sum_{\v{m}} p_{\v{m}} \ketbra{\psi_{\v m}} $ obtained by discarding the measurement record. 

The stability of entanglement in the volume-law phase ($p < p_c$) is explained qualitatively by the emergence of a quantum error correcting code that manages to hide information from future measurements for a long time~\cite{choi_quantum_2020,gullans_dynamical_2020,fan_self-organized_2021}. This coding perspective is illustrated concretely by the behavior of a reference qubit $R$ initially entangled with a monitored many-body system~\cite{gullans_scalable_2020}.
Let $S_R(t)$ be the entanglement entropy of the reference qubit as a function of time $t$ in the monitored dynamics; at late times one generally has $S_R(t) \sim e^{-t / \tau}$, with a time constant $\tau$ dubbed the {\it purification time}. The behavior of $\tau$ changes sharply at the transition. 
In the volume-law phase it obeys $\tau \sim \exp(N)$, signifying a successful encoding of at least some information about the state of $R$, which is protected from measurements for a long time. In the area-law phase $\tau$ becomes $O(1)$, showing that the encoding fails. Finally, at the critical point ($p = p_c$) $\tau$ diverges algebraically, $\tau \sim N^{z/d}$, with $z$ a dynamical critical exponent (typically $z = 1$~\cite{zabalo_operator_2022}) and $d$ the spatial dimension. 
In the following we will make use of the purification time as a practical diagnostic for the underlying entanglement phase, as it is numerically more efficient to compute than the entanglement entropy of large subsystems. 

In one dimension, the entanglement phase transition also corresponds to a transition in the classical simulation complexity of the dynamics: area-law states in 1D have constant entanglement, and can be efficiently simulated with MPS methods. This is the principle behind the SEBD algorithm~\cite{napp_efficient_2022}, in which a 2D sampling task is reduced to a $(1+1)$D monitored dynamics and simulated by MPS methods in the area-law phase.
Ref.~\cite{vovk_entanglement-optimal_2022} has extended this approach to continuous-time Markovian open-system dynamics: the system-environment coupling, in the form of a Lindbladian master equation, can be ``unraveled'' into trajectories~\cite{dalibard_wave-function_1992} (stochastic pure-state evolutions) that contain quantum measurements, which in turn can lower entanglement and help the accuracy of MPS simulation. As the unraveling is a simulation artifact and is not physical, it can be optimized to minimize the amount of entanglement. Ref.~\cite{vovk_entanglement-optimal_2022} proposes an adaptive scheme that chooses the unraveling with the lowest expected value of the entropy at each position and time during the evolution; above a threshold noise strength, the trajectories enter an area-law phase and efficient MPS simulation becomes possible. 

In this work we build on the approaches of Refs.~\cite{napp_efficient_2022, vovk_entanglement-optimal_2022} to address the problem of sampling from {\it noisy, shallow} circuits in 2D. 


\section{Unraveling noise into monitored trajectories \label{sec:unraveling}}

In this Section we review the basics noise models and their unraveling and discuss the entanglement-optimal unraveling to use in the noisy-SEBD algorithm. 

\subsection{Noise models and unraveling \label{sec:noise_models}}

In quantum computers, it is often a good approximation to treat the inevitable interactions with the environment as Markovian noise, modeled by quantum channels (completely-positive trace preserving maps on the space of density matrices~\cite{nielsen_quantum_2000}). 
Mathematically a quantum channel $\Phi$ can be represented by as a sum of Kraus operators $\{M_i\}$,
\begin{align}
	\Phi(\rho)=\sum_iM_i\rho M^\dagger_i,
\end{align}
which must obey the trace-preservation condition:
\begin{align}\label{eq:completeness}
	\sum_i M_i^\dagger M_i =\mathbb{I}.
\end{align}
As examples, the dephasing channel can be represented by
\begin{align}\label{eq:dephase}
	\Phi(\rho)  = (1-\varepsilon)\rho + \varepsilon Z\rho Z,
\end{align}
i.e. with Kraus operators $\{\sqrt{1-\varepsilon}\mathbb{I}, \sqrt{\varepsilon}Z\}$,
and the depolarizing channel by
\begin{align}\label{eq:depolarize}
	\Phi(\rho)  = (1-\varepsilon)\rho + \frac{\varepsilon}{3} \(X\rho X + Y\rho Y +Z\rho Z\),
\end{align}
with Kraus operators $\{\sqrt{1-\varepsilon}\mathbb{I}, \sqrt{\varepsilon/3} X,  \sqrt{\varepsilon/3} Y,  \sqrt{\varepsilon/3} Z\}$. In both cases $\varepsilon$ is the noise strength.

One important property of the Kraus operators representation is that it is not unique. For a given quantum channel, different sets of Kraus operators are equivalent under unitary transformations $U$:
\begin{align}\label{eq:unitaryequival}
	M'_j = \sum_i U_{ji}M_i.
\end{align}
This equivalence also holds for non-square, semi-unitary transformations $U$ that satisfy only $U^\dagger U=\mathbb{I}$.
As an important consequence, even when a channel $\Phi$ can be unraveled into unitary processes (such as the dephasing and depolarizing channels above), this equivalence allows the freedom to choose an unraveling into non-unitary operators, which correspond to measurements.  For example, the dephasing channel can be unraveled into a stochastic projective measurement,
\begin{align}
	M_0 = \sqrt{1-2\varepsilon}\mathbb{I},\ 
	M_1=\sqrt{2\varepsilon}\vert0\rangle\langle0\vert,\ 
	M_2=\sqrt{2\varepsilon}\vert1\rangle\langle1\vert 
	\label{eq:dephasing_to_projective}
\end{align}
(i.e., a projective measurement of $Z$ taking place with probability $2\varepsilon$).
It can also be unraveled into a weak measurement of $Z$,
\begin{equation}
	\begin{aligned}
	M_0 & = \sqrt{1-\varepsilon}\cos\theta\mathbb{I}-\sqrt{\varepsilon}\sin\theta Z, \\ 
	M_1 & = \sqrt{1-\varepsilon}\sin\theta\mathbb{I}+\sqrt{\varepsilon}\cos\theta Z,
	\end{aligned}
	\label{eq:dephasing_to_weak}
\end{equation}
where $\theta$ is a free parameter tuning the strength of the measurement ($\theta = 0$ returns a unitary unraveling).


\subsection{Sampling noisy circuits \label{sec:sampling}}

Returning to the noisy circuit problem, we wish to sample from the distribution 
$P_{\mathcal N}(\v{z}) = \bra{\v{z}} \mathcal{N}(\ketbra{\v{0}}) \ket{\v{z}}$, 
with $\mathcal{N}$ the channel describing the noisy circuit evolution:
\begin{equation}
	\mathcal{N} = \Phi^{\otimes N} \circ \mathcal{U}_T \circ \Phi^{\otimes N} \circ \mathcal{U}_{T-1} \circ \cdots \Phi^{\otimes N} \circ \mathcal{U}_1.
\end{equation}
Here $\Phi$ is a single-qubit noise of strength $\varepsilon$ as before, while $\mathcal{U}_t(\rho) = u_t \rho u_t^\dagger$ describes the $t$-th layer of the (ideal) shallow unitary circuit, $U = u_T u_{T-1} \cdots u_1$. 

By expanding each $\Phi$ into its Kraus operators, $\Phi(\rho) = \sum_i M_i \rho M_i^\dagger$, we can rewrite the probability distribution $P_{\mathcal N}(\v{z})$ as
\begin{align}
	P_{\mathcal N}(\v{z}) 
	& = \bra{\v{z}} \mathcal{N}(\ketbra{\v{0}}) \ket{\v{z}} = \sum_{\v{m}} | \bra{\v{z}} \mathbb{M}_{\mathbf m} \ket{\v{0}} |^2, \label{eq:joint_prob} \\
	\mathbb{M}_{\mathbf m} & = \left( \bigotimes_{x=1}^N M_{m_{T,x}} \cdot u_T \right) \cdots \left( \bigotimes_{x=1}^N M_{m_{1,x}} \cdot u_1 \right), \label{eq:bigM_kraus}
\end{align}
where each index $m_{t,x}$ labels the Kraus operator for the noise channel $\Phi$ acting on qubit $x$ at step $t$, and $\mathbf{m}$ is a shorthand for the whole collection of indices $\{m_{t,x}\}$. Thus the operators $\{ \mathbb{M}_{\mathbf m} \}$ are a set of Kraus operators for the channel $\mathcal{N}$.
We can view each $\v m$ as a {\it quantum trajectory} of the evolution: the initial state $\ket{\v 0}$ evolves into the pure state $\ket{\psi_{\v m}} \equiv \mathbb{M}_{\v m} \ket{\v 0} / \bra{\v 0} \mathbb{M}_{\v m}^\dagger \mathbb{M}_{\v m}\ket{\v 0}^{1/2}$ with probability $\bra{\v 0} \mathbb{M}_{\v m}^\dagger \mathbb{M}_{\v m}\ket{\v 0}$; the true (mixed) state $\mathcal{N}(\ketbra{\v 0})$ is recovered as a stochastic mixture of the trajectories. 
It is straightforward to see that $|\bra{\v z} \mathbb{M}_{\v m} \ket{\v 0}|^2 \equiv P_{\mathcal N}(\v{z}, \v{m})$ is the {\it joint} probability of drawing trajectory $\v m$ and sampling bitstring $\v z$ at the end. 
Then, Eq.~\eqref{eq:joint_prob} can be written as 
\begin{equation}
	P_{\mathcal N}(\v z) = \sum_{\v m} P_{\mathcal N} (\v{z}, \v{m}),
\end{equation}
i.e. the marginal distribution obtained by summing over trajectories. 

This insight is widely used in simulations of open-system dynamics~\cite{dum_monte_1992,dalibard_wave-function_1992,gardiner_wave-function_1992,tian_quantum_1992}. The trajectory method allows one to simulate pure states rather than density matrices, which is often much more memory-efficient. This comes at the expense of having to average over many trajectories, e.g. in order to Monte Carlo-sample the expectation value of a target operator.
In our case, however, we only aim to {\it sample} from the distribution $P_{\mathcal N}(\v z)$, so there is no need for trajectory averaging: any joint sample $(\v{z},\v{m})$ drawn from a simulation of the trajectory dynamics yields a valid sample $\v{z}$ from the desired distribution $P_{\mathcal N}(\v z)$.
We emphasize that the $\v{m}$ samples and their distribution are purely mathematical artifacts of the method: the decomposition of $\Phi$ into Kraus operators includes a gauge degree of freedom that can be fixed arbitrarily. However, the marginal distribution $P_{\mathcal{N}}(\v{z})$ is gauge-invariant and physical, corresponding to the true experimental distribution of bitstrings. 

Beyond the practical advantage of simulating pure rather than mixed state evolution, the gauge degree of freedom in the choice of unraveling can be exploited to minimize entanglement within the trajectory~\cite{vovk_entanglement-optimal_2022}, thus extending the applicability of tensor network methods. Below we address the question of which unraveling yields the lowest entanglement for a given channel.



\subsection{Entanglement-optimal unravelings}\label{sec:optimalunravel}

For a given model of dynamics (e.g. a Hamiltonian or an individual instance of a RUC), one can locally optimize the unraveling of each noise channel $\Phi$ separately~\cite{vovk_entanglement-optimal_2022}. 
Here we take a simpler approach, and look for a general prescription that works well {\it on average} over RUCs. 
Specifically, we aim to exploit the area-law phase in monitored dynamics (Sec.~\ref{sec:mipt_review}), which is driven by the density of measurements in the dynamics. Thus we look for an unraveling of $\Phi$ into measurements, so as to increase the effective density of measurements and facilitate a transition to the area-law phase~\cite{kolodrubetz_optimality_2023}. As we have seen in Sec.~\ref{sec:noise_models}, there are multiple inequivalent ways of unraveling noise into measurements (e.g. weak vs stochastic projective measurements);
therefore we look for the unraveling $\v{M} = \{M_i\}$ of the nose channel $\Phi$ that has the strongest disentangling effect on the dynamics.

Working at a ``mean-field'' level in Haar-random circuits, we consider the scaling of average purity\footnote{Note that the purity $\mathcal{P} = \Tr(\rho^2)$ is related to the second Renyi entropy $S_2$ via $\mathcal{P} = e^{-S_2}$. Using convexity and the fact that $S_2 \leq S_\alpha$ for all $\alpha < 2$, we have a lower bound $\overline{S_\alpha} \geq -\ln \overline{P}$ between the average purity and the average Renyi entropies $S_{\alpha <1}$. The latter determine compressibility of the state into MPS form~\cite{schuch_entropy_2008}.}. in trajectories of the dynamics within a two-replica setting. This maps onto the partition function of a $\mathbb{Z}_2$ Ising magnet, whose ordered and disordered phase corresponds to volume-law and area-law entanglement, respectively. Our goal is to facilitate simulation by minimizing entanglement, therefore we aim to {\it minimize} the couplings in the magnet.
The problem is analyzed in Appendix~\ref{ap: statmech}, where we show that the minimization of the coupling amounts to maximizing the following objective function:
\begin{align}\label{eq:targetfunction}
	x(\v{M}) = \sum_i\frac{\tr\(M_i^\dagger M_iM_i^\dagger M_i\)}{2\tr\(M_i^\dagger M_i\)}.
\end{align}
This has a natural physical interpretation: if we view the Kraus operators $\{M_i\}$ as {\it instruments} of a generalized measurement (positive operator-valued measure) $\{M_i^\dagger M_i\}$, and apply such measurement to the fully mixed state $\mathbb{I}/2$, we obtain post-measurement states $\rho_i \equiv M_i M_i^\dagger / \Tr(M_i M_i^\dagger)$ with probabilities $\pi_i \equiv \Tr(M_i^\dagger M_i)/2$; the objective function is given by the average purity of the post-measurement states: $x(\v{M}) = \sum_i \pi_i \Tr(\rho_i^2)$.
Thus within this approach, the unraveling that minimizes many-body entanglement in the RUC dynamics is also the one that best purifies a single mixed qubit.
Also this is a particular feature of Haar-random circuits and likely not true of more structured models, since under Haar-random unitaries $\mathbb{I}/2$ is the natural averaged reduced density matrix for a single qubit.

We also note that this cost function (post-measurement purity of a mixed qubit) is in fact identical to the entanglement of a Bell pair state after measuring one qubit, which was proposed in Ref.~\cite{kolodrubetz_optimality_2023} as a heuristic measure of the disentangling power of different unravelings. 


The optimal unraveling $\v{M}_{\rm opt}$ is given by
\begin{align}
	\v{M}_{\rm opt} = {\rm argmax}_{\v{M}}\[x(\v{M})\],
\end{align}
where $\v{M} = \{M_i\}$ ranges over Kraus decompositions of $\Phi$, and thus is subject to (semi-)unitary gauge freedom per Eq.~\eqref{eq:unitaryequival}. 
In general, the semi-unitary gauge freedom makes the optimization nontrivial as it allows for infinitely many parameters. However, below we show that for a broad and physically relevant class of quantum channels there exists an optimal unraveling of minimal rank which can be obtained analytically.

\begin{figure*}
	\centering
	\includegraphics[width=\textwidth]{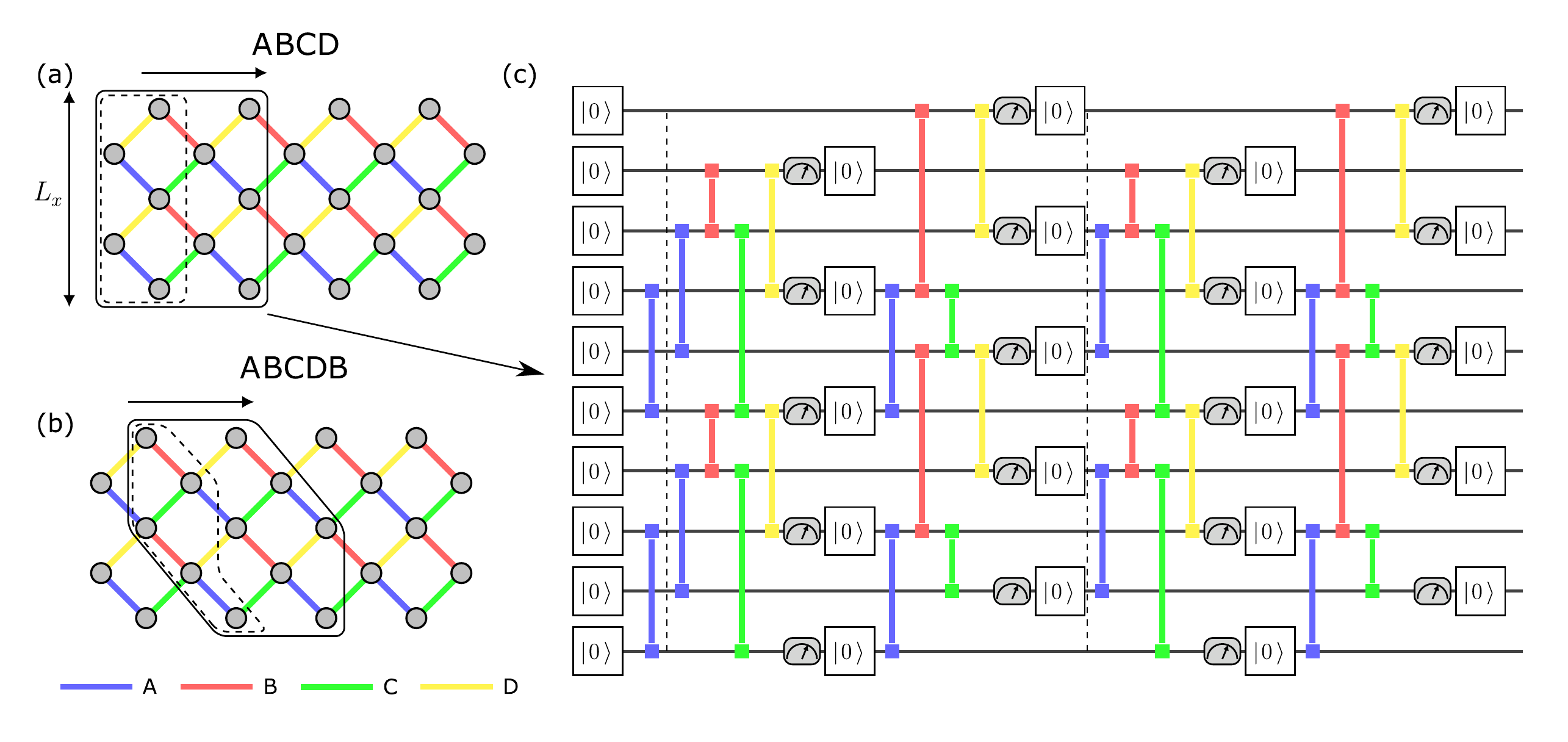}
	\caption{Schematic depiction of the 2D qubit array and effective 1D subsystem used in the SEBD algorithm. 
	(a,b) Example of 2D qubit array with $L_x=5$, showing the gate sequences used: ABCD for depth $T = 4$ (a), ABCDB for depth $T=5$ (b).
	The region enclosed by the solid loop corresponds to the effective 1D subsystem, and the region enclosed by the dashed loop corresponds to sites which are measured after applying gates in the past lightcone. (c) Equivalent 1D circuit for $T=4$ and $L_x=5$ with gate sequence ABCD. The effective 1D system has size $2L_x=10$, with gates up to the third nearest neighbor, and each measurement is followed by a reset to the $\vert0\rangle$ state. The dashed lines enclose a unit cell whose architecture repeats periodically in time.}
	\label{fig:grid}
\end{figure*}


\subsection{Unital Qubit Channels}\label{sec:unital}

Let us consider {\it unital} qubit channels, which are defined as those that leave the fully mixed state $\mathbb{I}/2$ invariant.
Up to unitary transformations (which we can ignore in the RUC setting), unital qubit channels have the canonical form~\cite{li_unital_2023}
\begin{align}\label{eq:unitalchannel}
	\Phi(\rho)=p_0\rho+p_x X\rho X+p_y Y\rho Y +p_z Z\rho Z,
\end{align}
with $p_\alpha \geq 0$ and $\sum_\alpha p_\alpha = 1$. 
We start with a set of $n$ Kraus operators $\v{M} = \{M_i\}$, where each element can be represented as
\begin{align}\label{eq:Krausoperatoransatz}
	M_i = a_i\mathbb{I}+b_i\v{\sigma}\cdot\tilde{\v{u}}_i, 
\end{align} 
where $a_i$ and $b_i$ are real non-negative numbers\footnote{Note we can assume $a_i$ real and non-negative by redefining $M_i$ by an overall phase; the phase of $b_i$ can then be absorbed into $\tilde{\v u}_i$.}, $\tilde{\v{u}}_i=\(e^{i\phi_{i,x}}u_{i,x},e^{i\phi_{i,y}}u_{i,y},e^{i\phi_{i,z}}u_{i,z}\)$ are complex unit vectors (i.e. $\tilde{\v u}^\ast_i \cdot \tilde{\v u}_i = 1$), and $\v{\sigma}=\(X, Y, Z\)$. 
The Kraus operators of Eq.~\eqref{eq:Krausoperatoransatz} must yield the Choi matrix of the unital channel $\Phi$ in Eq.~\eqref{eq:unitalchannel}, i.e. they must satisfy
\begin{equation}
	\sum_i M_i \otimes M_i^\ast = \sum_{\alpha = 0,x,y,z} p_\alpha \sigma^\alpha \otimes (\sigma^\alpha)^\ast
\end{equation}
(we set $\sigma^0 = \mathbb{I}$). 
This gives the following four relations:
\begin{align}
	\sum_i a_i^2 = & p_0,\quad\sum_i b_i^2 = 1-p_0, \label{eq:abconstrain}\\
	\sum_i a_ib_i\tilde{\v{u}}_i = & \v{0},\quad
	\sum_i \tilde{u}^\ast_{i,\alpha} \tilde{u}_{i,\beta} b_i^2= p_\alpha\delta_{\alpha\beta}\label{eq:uconstrain}.
\end{align}

The optimization of the target function $x(\v{M})$ subject to these constraints is carried out in Appendix~\ref{ap:inequality}.
The objective function is maximized when $a_i = \sqrt{p_0/n}$ and $b_i = \sqrt{(1-p_0)/n}$ for all $i$, and the unit vectors $\tilde{\v{u}}_i$ are real: $\tilde{\v u}_i = \v{u}_i = (u_{i,x}, u_{i,y}, u_{i,z})$.
The  remaining constraints on the vectors $\{ \v{u}_i \}$, Eq.~\eqref{eq:uconstrain}, read
\begin{align}
	\sum_i\v{u}_i=\v{0},
	\quad
	\sum_iu_{i,\alpha}u_{i,\beta} = \frac{n p_\alpha}{1-p_0} \delta_{\alpha\beta}.
	\label{eq:deformed_2des}
\end{align}
As an example, a solution with $n=4$ is
\begin{equation}
	\left\{
	\begin{aligned}
		\v{u}_0 & =\(\sqrt{p_x}, \sqrt{p_y}, \sqrt{p_z} \) / \sqrt{1-p_0},\\
		\v{u}_1 & =\(\sqrt{p_x}, -\sqrt{p_y}, -\sqrt{p_z} \) / \sqrt{1-p_0},\\
		\v{u}_2 & =\(-\sqrt{p_x}, \sqrt{p_y}, -\sqrt{p_z} \) / \sqrt{1-p_0},\\
		\v{u}_3 & =\(-\sqrt{p_x}, -\sqrt{p_y}, \sqrt{p_z} \) / \sqrt{1-p_0},	
	\end{aligned}
	\right.
	\label{eq:deformed_tetrahedron}
\end{equation}
i.e. the vertices of a regular tetrahedron inscribed in the Bloch sphere, up to a rescaling $\sqrt{3p_\alpha / (1-p_0)}$ of each axis.

In the case of $p_x = p_y = p_z$ (depolarizing noise, Eq.~\eqref{eq:depolarize}), the conditions in Eq.~\eqref{eq:deformed_2des} become 
\begin{equation}
	\mathbb{E}_i[ \v{u}_i] = \v{0},
	\qquad 
	\mathbb{E}_i[u_{i,\alpha}u_{i,\beta}] = \frac{1}{3} \delta_{\alpha \beta},
	\label{eq:spherical_design}
\end{equation}
where $\mathbb{E}_i$ denotes averaging over $i$ with respect to the uniform probability distribution $\textsf{Pr}(i) = 1/n$. 
The conditions in Eq.~\eqref{eq:spherical_design} define a {\it spherical 2-design}, i.e. a probability distribution on the sphere whose first two moments coincide with those of the uniform distribution. Such distributions exist if $n = 4$ or $n \geq 6$. In particular, the minimal ($n = 4$) optimal unraveling is 
\begin{equation}
	\label{eq:weakmeasurementsdepolarizing}
	\left\{ 
	\begin{aligned}
	M_0&=\sqrt{\frac{1-\varepsilon}{4}}\mathbb{I}+\sqrt{\frac{\varepsilon}{12}}\(X +Y+Z\),\\
	M_1&=\sqrt{\frac{1-\varepsilon}{4}}\mathbb{I}+\sqrt{\frac{\varepsilon}{12}}\(X -Y-Z\),\\
	M_2&=\sqrt{\frac{1-\varepsilon}{4}}\mathbb{I}+\sqrt{\frac{\varepsilon}{12}}\(-X +Y-Z\),\\
	M_3&=\sqrt{\frac{1-\varepsilon}{4}}\mathbb{I}+\sqrt{\frac{\varepsilon}{12}}\(-X -Y+Z\),
	\end{aligned}
	\right.
\end{equation}	
i.e., a weak measurement along 4 directions corresponding to the vertices of a regular tetrahedron. 
Similarly, a solution to Eq.~\eqref{eq:spherical_design} with $n = 6$ is given by the vertices of a regular octahedron, corresponding to weak measurements of the Pauli $X$, $Y$ and $Z$ operators. 

For the dephasing channel Eq.~\eqref{eq:dephase} ($n=2$), an optimal unraveling is
\begin{equation}
	\label{eq:weakmeasurementsdephasing}
	\left\{ 
	\begin{aligned}
	M_0 & = \sqrt{\frac{1-\varepsilon}{2}}\mathbb{I}+\sqrt{\frac{\varepsilon}{2}}Z, \\
	M_1 & = \sqrt{\frac{1-\varepsilon}{2}}\mathbb{I}-\sqrt{\frac{\varepsilon}{2}}Z.
	\end{aligned}
	\right.
\end{equation}
In all these cases, the most disentangling unraveling takes the form of {\it weak measurements} rather than stochastic projective measurements (e.g. Eq.~\eqref{eq:dephasing_to_projective} for the case of dephasing). 

In Appendix~\ref{ap:1Doptimal} we verify that, for the optimal unraveling Eq.~\eqref{eq:weakmeasurementsdephasing}, the measurement-induced phase transition occurs at a lower value of $\varepsilon$ ($\varepsilon_c \simeq 0.044$) compared to the unraveling into stochastic projective measurements Eq.~\eqref{eq:dephasing_to_projective} ($\varepsilon_c \simeq 0.084$---note the probability of doing a measurement is $p = 2\varepsilon$ and the well-known MIPT critical point is at $p_c\simeq 0.168$~\cite{zabalo_critical_2020}). 
We see that our simple mean-field approach is already sufficient to lower the critical noise strength by almost a factor of 2.
It would be interesting to test whether locally- and adaptively-optimized unravelings as in Ref.~\cite{vovk_entanglement-optimal_2022} could further improve this threshold.
For the rest of this work, we will use the optimal unraveling in Eq.~\eqref{eq:weakmeasurementsdepolarizing} and Eq.~\eqref{eq:weakmeasurementsdephasing} for depolarizing and dephasing noise channels unless otherwise specified.

Before moving on, we note that, although the above solution works only for unital channels, other physically relevant models may also be analytically tractable. In particular it is straightforward to find an optimized unraveling for the amplitude damping channel, a paradigmatic non-unital channel defined by Kraus operators
\begin{align}
		\left\{ \begin{aligned}
			M_0=&\begin{pmatrix}
				1&0\\
				0&\sqrt{1-\varepsilon}
			\end{pmatrix},\\
			M_1 =& \begin{pmatrix}
				0&\sqrt{\varepsilon}\\
				0&0
			\end{pmatrix}.
		\end{aligned}\right.
\end{align}
In Appendix \ref{ap:1Ddamp} we consider the optimization over $2\times2$ unitary rotations of these Kraus operators and derive the optimized unraveling
\begin{align}
	\left\{ \begin{aligned}
		{M}_0=&\begin{pmatrix}
			1&\sqrt{\varepsilon}\\
			0&\sqrt{1-\varepsilon}
		\end{pmatrix}/\sqrt{2},\\
		{M}_1 =& \begin{pmatrix}
			-1&\sqrt{\varepsilon}\\
			0&-\sqrt{1-\varepsilon}
		\end{pmatrix}/\sqrt{2}.
	\end{aligned}\right.
\end{align}	
Furthermore, we numerically verify the disentangling effect of this unraveling by studying the critical noise strength of the measurement-induced phase transition: $\varepsilon_c\approx0.17$ for the optimized unraveling, compared with $\varepsilon_c\approx0.29$ for the original unraveling, again almost a factor of 2 improvement.
Finally, beyond analytically-tractable cases, one can always perform the optimization numerically for arbitrary single-qubit noise models, by restricting to a finite number of Kraus operators.


\section{Noisy-SEBD algorithm \label{sec:nsebd}}

\subsection{Description of the algorithm}

We consider noisy random circuits in 2D with finite depth $T$ acting on a grid with $N=L_x \times L_y$ qubits. The goal of our algorithm is to sample bitstrings $\v{z}$ from the distribution $P_{\mathcal N}(\v z)$, Eq.~\eqref{eq:joint_prob}, determined by the circuit instance and noise channel $\Phi$ (whose parameters we assume are known), with a small error. 

Let us first review the noiseless case, studied in Ref.~\cite{napp_efficient_2022} and illustrated in Fig.~\ref{fig:idea}(a).
Due to locality, the outcome $z_i$ on any given qubit only depends on the evolution within its past lightcone. Thus to sample all the outcomes $z_i$ on the first row of qubits, $y = 1$, we only need to apply gates and channels on qubits within the past lightcone of the line $\{(x,y=1,t=T)\}$, which includes qubits with $y \leq T$. This corresponds to a circuit of depth $T$ on $\leq L_x T$ qubits, which can be simulated efficiently via MPS methods. 
At this point one can successfully sample the outcomes $z_i$ for the first row of qubits, and move on to the second row ($y = 2$) iterating the same approach, performing only the gates and channels within the past lightcone of $\{(x,y=2,t=T)\}$, etc. 
This effectively maps the 2D shallow circuit to an equivalent 1D circuit, where the readout measurements are converted to mid-circuit measurements and resets (See Fig.~\ref{fig:grid}(c) as an example). As a result of this mapping, the spatial direction along $L_y$ becomes the time direction for the 1D circuit (an idea also known as {\it space-time duality} in quantum circuits~\cite{ippoliti_postselection-free_2021,lu_spacetime_2021,ippoliti_fractal_2022}).
The SEBD algorithm is based on MPS simulation of this effective 1D dynamics for the purpose of sampling the $\v{z}$ outcomes.

The issue with iterating this approach indefinitely is that the entanglement in the quasi-1D state of $L_x \times T$ qubits in principle grows with each step, up to the point where MPS simulation fails. 
However, Ref.~\cite{napp_efficient_2022} observed that, due to the mid-circuit measurements, the effective dynamics may enter an area-law phase, wherein the entanglement remains finite and approximate sampling can be carried out efficiently with high accuracy. 
In general, the effective 1D circuit consists of $N_{\rm 1D}=cT L_x$ qubits, where $c$ is a constant depending on the lattice geometry (the slope of the lightcone, in the models considered here $c\approx 1/2$), and the spatial range of the two-qubit gates is proportional to $T$. Equivalently, one may view the effective system as a quasi-1D strip of size $L_x \times cT$ with nearest-neighbor gates in both directions.
Either way, each circuit layer on $cTL_x$ qubits is followed by the measurement of $L_x$ qubits (a full row), giving a ratio of measurements to unitary operation of $\sim 1/T$. When this ratio is sufficiently high (i.e. $T$ sufficiently low), the system enters an area-law phase and SEBD is efficient. 

Let us now add noise to the picture. As discussed in Sec.~\ref{sec:sampling}, we can unravel the noise channels $\Phi$ into an arbitrary set of Kraus operators, simulate the pure-state trajectories, sample from the joint distribution $P_{\mathcal N}(\v{z}, \v{m})$ and keep only the $\v{z}$ samples. 
We adopt the entanglement-optimal unraveling of unital noise channels discussed in Sec.~\ref{sec:optimalunravel} to suppress entanglement in the effective 1D state at the level of quantum trajectories. 
The simulated dynamics now features mid-circuit measurements with two distinct origins: a density $\propto 1/T$ coming from the final sampling step in the 2D circuit, and a density $\propto \varepsilon$ coming from the unraveling of noise. 
Below a critical noise rate $\varepsilon_c$, dependent on the model and the circuit depth $T$, the entanglement of effective 1D state satisfies area law, and thus allows efficient classical simulation. 
These ideas are schematically summarized in Fig.~\ref{fig:idea}.

\begin{figure}
	\centering
	\includegraphics[width=0.5\textwidth]{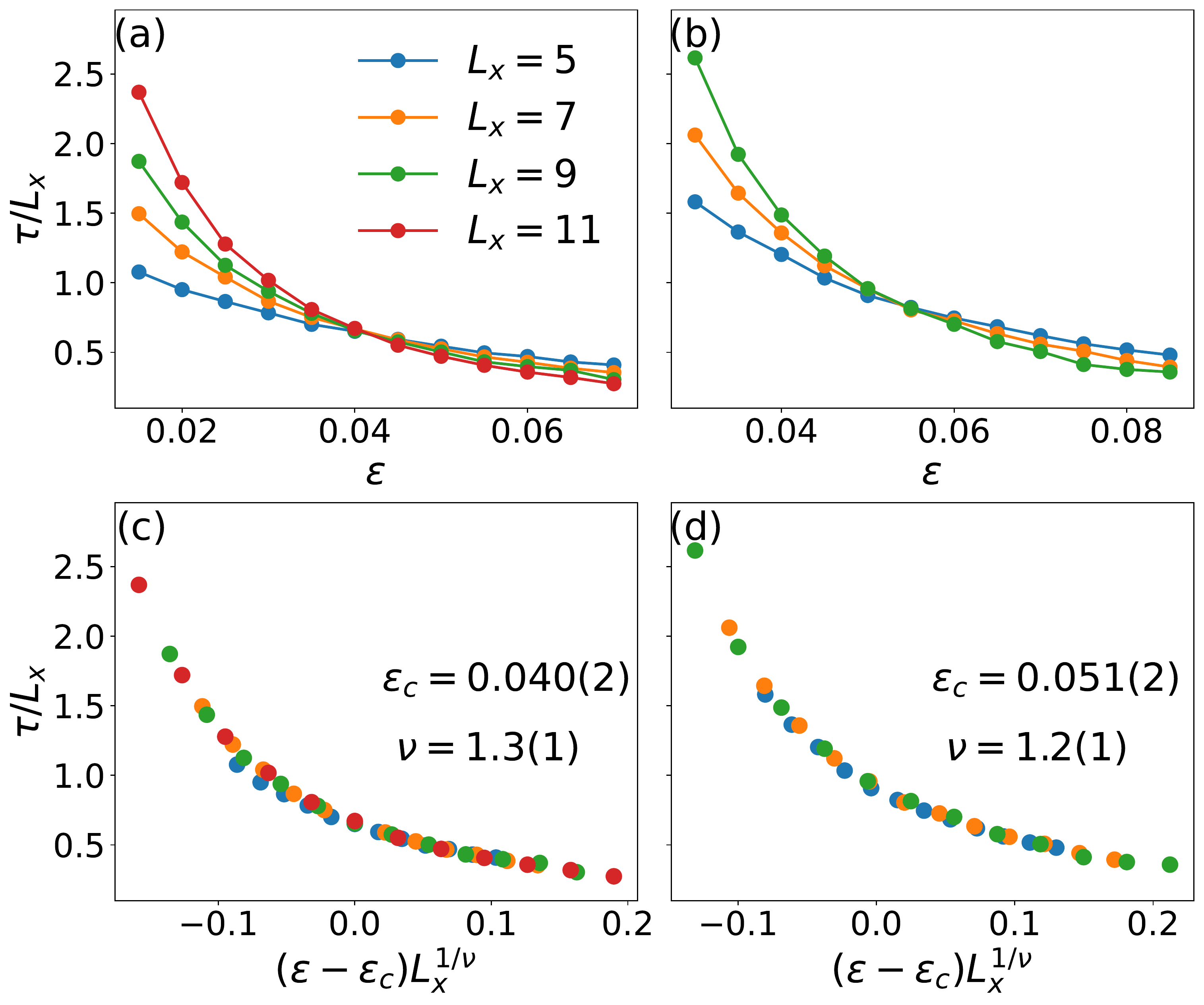}
	\caption{Purification time $\tau$ of a single probe qubit as a function of the noise rate $\varepsilon$ for (a) depth $T=4$ with gate sequence ABCD and (b) depth $T=5$ with gate sequence ABCDB. Scaling collapse of the data for (c) $T=4$ and (d) $T=5$. The data are averaged over $5\cdot10^3-3\cdot10^4$ realizations.
	}
	\label{fig:PT}
\end{figure}

\subsection{Numerical results: entanglement phase transition}\label{sec:numerics}

In this section, we numerically study the entanglement phase transition in the effective 1D subsystem that is used in the noisy-SEBD algorithm (Fig.~\ref{fig:grid}(c)). 
As an example, we choose shallow circuits that act on a 2D square lattice (Fig.~\ref{fig:grid}(a-b)) with unitary gates similar to those employed in Google's RCS experiment~\cite{arute_quantum_2019}.
The two-qubit gates are iSWAP-like fermionic simulation gates 
\begin{align}
	{\rm fSim}(\pi/2, \pi/6)=\begin{pmatrix}
		1&0&0&0\\
		0&0&-i&0\\
		0&-i&0&0\\
		0&0&0&e^{-i\pi/6}
	\end{pmatrix},
\end{align}
sandwiched between single-qubit rotations randomly chosen from the set $\{ X^{\pm 1/2}, Y^{\pm 1/2}, W^{\pm 1/2}, V^{\pm 1/2} \}$, where $W=(X+Y)/\sqrt{2}$ and $V=(X-Y)/\sqrt{2}$ are Hadamard-like gates (note that $W^{\pm 1/2}$ and $V^{\pm 1/2}$ are non-Clifford).
The two-qubit gates are applied to bonds of the square lattice according to the sequence ABCD for $T=4$ (Fig.~\ref{fig:grid}(a)) and ABCDB for $T=5$ (Fig.~\ref{fig:grid}(b)). These specific sequences are chosen to be the most entangling for the effective 1D dynamics, defined as giving the longest single-qubit purification time as discussed in the following. The noise channel $\Phi$ is taken to be the depolarizing channel, Eq.~\eqref{eq:depolarize}, with strength $\varepsilon$.

As a diagnostic for the entanglement phase transition which underpins the efficiency of noisy-SEBD, we use the single-qubit purification time $\tau$ discussed in Sec.~\ref{sec:mipt_review}.
This is advantageous from the numerical point of view, relative to a direct calculation of the bipartite entanglement entropy, since it does not suffer from the large finite-size drifts arising from the logarithmic divergence of entropy at the critical point~\cite{zabalo_critical_2020}. 
The phase transition between area-law and volume-law entanglement scaling is probed by the mutual information between the single reference qubit and the rest of the system. In practice, we introduce a reference qubit which initially forms a Bell pair with a system qubit. At later time the average entropy of the reference qubit is captured by the exponential decaying relation $S_R(t)\sim e^{-t/\tau}$. In the volume-law phase (low measurement/noise rate), $S_R$ can remain nonzero for a long time $\tau\sim\exp(L_x)$. 
Physically this implies that there is finite measurement-induced entanglement between opposite ends of the system as one takes $L_x$, $L_y$ to infinity jointly with $L_y = {\rm poly}(L_x)$; this can be interpreted as emergent quantum teleportation~\cite{bao_finite_2022} and has recently been explored experimentally~\cite{hoke_measurement-induced_2023}. 
On the other hand, in the area-law phase (high measurement/noise rate), $S_R$ decays rapidly to zero with $\tau = O(1)$. At the critical point, we expect $\tau\sim L_x^z$, where $z$ is the dynamical critical exponent (typically $z = 1$ for measurement-induced transitions in short-range interacting 1D systems~\cite{zabalo_critical_2020, gullans_scalable_2020}). 

In Fig.~\ref{fig:PT}, we show a finite size scaling analysis of $\tau/L_x$ for both $T=4$ (gate sequence ABCD) and $T=5$ (gate sequence ABCDB), obtained from MPS simulation of the space-wise dynamics up to $L_y = 2L_x$, where only $L_x\leq L_y \leq 2L_x$ are used for fitting to avoid the early-time transient effect. The existence of a finite-size crossing point of the ratio $\tau/L_x$ for all system sizes (Fig.~\ref{fig:PT}(a,b)) indicates $z=1$, which is consistent with the emergence of 1+1D conformal symmetry at the transition. Therefore, we use the scaling ansatz 
\begin{align}
	\tau(\varepsilon, L_x) = L_x F\[\(\varepsilon-\varepsilon_c\)L_x^{1/\nu}\],
\end{align}
to determine the location of the critical point $\varepsilon_c$ and the correlation length critical exponent $\nu$. 
From the data collapse, Fig.~\ref{fig:PT}(b)(d) (See Appendix~\ref{ap:collapse} for method details), we locate $\varepsilon_c=0.040(2)$ with $\nu=1.3(1)$ for $T=4$ (ABCD) and $\varepsilon_c=0.053(2)$ with $\nu=1.2(1)$ for $T=5$ (ABCDB). For both values of $T$ we find correlation length exponent $\nu\approx1.3$, which is consistent with the know value for the measurement-induced phase transition in 1D, as expected.
Moreover we see that increasing circuit depth $T$ leads to a larger critical noise rate $\varepsilon_c$, which is the qualitative behavior sketched in Fig.~\ref{fig:idea}(d). 

These numerical results support our qualitative expectations for the complexity of noisy-SEBD. Since the hardness of the method is exponential in $T$, obtaining accurate predictions for the phase boundary at larger $T$ becomes increasingly challenging. However, it is reasonable to conjecture that, for large $T$,
\begin{enumerate}
\item[(i)]  as the quasi-1D system of $cT\times L_x$ qubits approaches a 2D limit, one should recover the 2D measurement-induced phase transition which occurs at a finite noise rate $\varepsilon_{c,2D}$~\cite{turkeshi_measurement-induced_2020,sierant_measurement-induced_2022};
\item[(ii)] the transition should occur at a finite {\it total} noise rate, comprising the unraveled measurements (rate $\varepsilon$) and the sampling of the final state (rate $\sim 1/T$), thus $\varepsilon_c(T) = \varepsilon_{c,2D} + O(1/T)$.
\end{enumerate}
We find evidence in support of these conjectures in Clifford circuits, which can be simulated in polynomial time by the stabilizer method~\cite{aaronson_improved_2004} (though note that this limits us to the projective unraveling of the noise channel), see Appendix~\ref{ap:largeT_limit}.

Finally, in Appendix~\ref{ap:benchmark} we verify the accuracy of the noisy-SEBD algorithm by benchmarking its output against direct MPO simulations of the noisy dynamics and against stabilizer simulations of Clifford circuits. 

\begin{figure*}
	\centering
	\includegraphics[width=\textwidth]{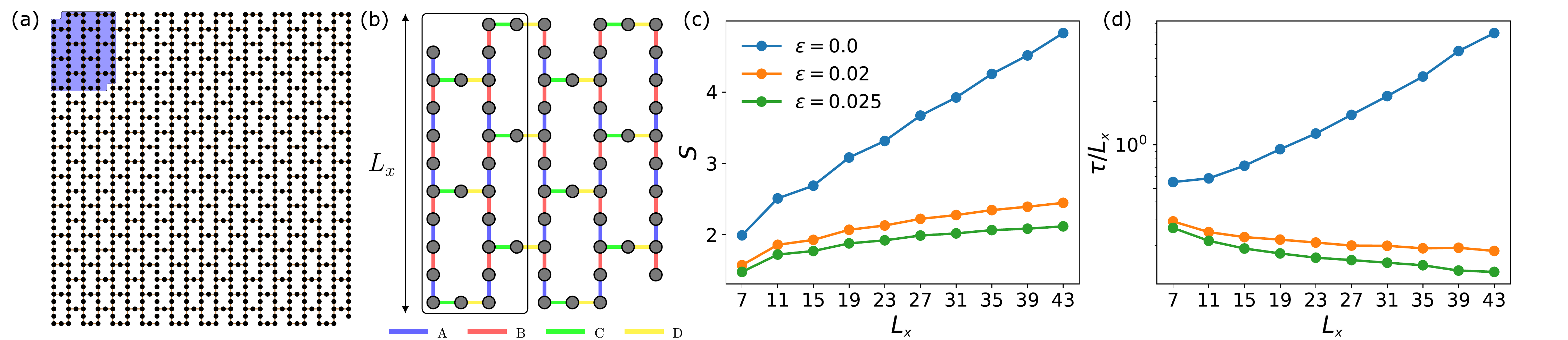}
	\caption{(a) Layout of 1121-qubit IBM quantum processor Condor;  65-qubit Hummingbird is shown as blue region.
	(b) Gate sequence for circuits with depth $T=5$ (ABCDA) on 65-qubit quantum processor Hummingbird, of linear size $L_x = 11$. Other IBM processors have similar layout: Eagle with $L_x = 15$, Osprey with $L_x = 27$, and Condor with $L_x = 43$. The region surrounded by a solid line depicts the 1D effective subsystem for noisySEBD simulation of random circuits with $T=5$ (gate sequence ABCDA). 
	(c) Bipartite entanglement entropy $S$ and (d) purification time $\tau$ of a reference qubit, for noiseless ($\varepsilon = 0$) and noisy ($\varepsilon = 0.02$ and $0.025$) circuits of depth $T=5$ as a function of linear system size $L_x$, which refers to subsets of the heavy-hexagon lattice in panel (a).
	The data are averaged over $10^3-10^4$ realizations of the random circuits.}
	\label{fig:ibm}
\end{figure*}

\subsection{Application: IBM quantum processors}\label{sec:ibm}

Quantitatively, the results in the previous section show that circuits on square lattice architectures, on NISQ platforms such as Google's Sycamore processor with native $\textsf{iSWAP}$-like gates and $\gtrsim 98\%$ two-qubit gate fidelities (translating to $\varepsilon \lesssim 0.01$ in our parametrization, see Appendix~\ref{ap:fidelity}), are already in the ``hard phase'' at depth $T = 4$. Since depth $T = 3$ is in the easy phase already for noiseless SEBD (i.e., for $\varepsilon = 0$), the inclusion of noise does not allow efficient simulation of an additional gate layer in this setting. 
This class of circuits is however a worst-case scenario, representing highly scrambling dynamics optimized for hardness of simulation~\cite{arute_quantum_2019,mi_information_2021}. 


Here we consider the application of noisy-SEBD to circuits on quantum processors with heavy-hexagon geometry and native CNOT gates, such as IBM Quantum's family of processors, Fig.~\ref{fig:ibm}(a,b).
We consider again highly-scrambling random ciruits with iSWAP two-qubit gates and single-qubit gates randomly chosen from $\{X^{\pm 1/2}, Y^{\pm1/2}, W^{\pm 1/2}, V^{\pm 1/2}\}$. An iSWAP gate can be compiled into two native CNOT gates, plus single-qubit gates. Thus the effective noise rate for iSWAP gates is twice the CNOT error, giving e.g. $\approx 96\%$ median gate fidelity on the Osprey processor (corresponding to $\varepsilon \approx 0.025$ in our parametrization, see Appendix~\ref{ap:fidelity}).

We consider subsystems of a qubit array based on the Condor processor, Fig.~\ref{fig:ibm}(a). The full system has $N = 1,121$ qubits and linear size $L_x = 43$. The gate sequence and unit cell for the SEBD algorithm are shown in Fig.~\ref{fig:ibm}(b), for the $N = 65$, $L_x = 11$ Hummingbird processor.
We characterize the efficiency of the {(noisy-)SEBD} algorithm by computing the half-system bipartite entanglement entropy $S$, Fig.~\ref{fig:ibm}(c), and single-qubit purification time $\tau$, Fig.~\ref{fig:ibm}(d), for both noiseless and noisy random circuits with depth $T=5$ (measured in units of iSWAP gates, thus corresponding to 10 CNOT gates on each qubit). 
As the linear system size is varied between $L_x = 7$ and $L_x = 43$, we observe a clear volume-law phase in the noiseless case, with volume-law entropy $S \propto L_x$ and exponential purification time $\tau \sim \exp(L_x)$. On the contrary, for noise rate $\varepsilon = 0.02$, we see a weak growth of $S$ with $L_x$, consistent with critical scaling $S \sim \ln(L_x)$ or eventual saturation to an area-law. The ratio $\tau / L_x$ also decreases, either to a finite constant (critical behavior) or to zero (area-law behavior). 
Finally, for $\varepsilon = 0.025$, both diagnostics are indicative of an area-law phase.

We conclude that in this case the inclusion of a realistic noise rate in the simulation algorithm is sufficient to drive a transition in complexity of the SEBD algorithm. This makes the difference between an asymptotically-efficient and asymptotically-hard MPS simulation of the sampling problem. 
We note that, while circuits with low depth such as $T = 5$ can likely be simulated by brute-force tensor network methods up to hundreds or thousands of qubits, the cost of such methods remains generically exponential in the linear size of the system $L_x$. On next-generation processors with $N\sim 10^5$ qubits those methods would become intractable, while noisy-SEBD would remain practical in the easy phase.


\section{Discussion \label{sec:discussion}}

We have introduced a classical algorithm, noisy-SEBD, to sample from the output distribution of noisy, shallow circuits in two dimensions. 
The algorithm uses the insight of mapping a 2D RCS problem to a 1D monitored dynamics problem (space-evolving block decimation, SEBD~\cite{napp_efficient_2022}), while also unraveling the action of noise into additional measurements on the system. At sufficiently low depth and sufficiently strong noise, this enables efficient MPS simulation of the monitored quantum trajectories, and thus efficient sampling from the appropriate noisy output distribution.

Given that the unraveling of noise into measurements is arbitrary, it can be optimized so as to reduce the amount of entanglement~\cite{vovk_entanglement-optimal_2022}. Here we have focused on a ``mean-field'' approach where single-qubit noise channels at all positions and times are unraveled in the same way, chosen based on the two-replica statistical mechanics description of the circuit upon averaging over random gates. We have found that, for unital channels (such as dephasing and depolarizing), the optimal unraveling is based on uniform weak measurements, rather than stochastic projective measurements. 
The difference between unravelings is substantial---in the standard model of brickwork circuits in 1D, the noise threshold corresponding to the measurement-induced entanglement transition is reduced by a factor of about 2 for the optimal weak-measurement unraveling ($\varepsilon_c\approx 0.04$) compared to the usual projective measurement unraveling ($\varepsilon_c\approx 0.08$, i.e. measurement rate $p_c = 2\varepsilon_c \approx 0.16$).
This is consistent with prior observations in Ref.~\cite{kolodrubetz_optimality_2023}, and the optimization technique could be of independent interest for the study of measurement-induced entanglement transitions. 

While noisy RCS in the anticoncentration regime was shown to be classically-simulable in polynomial time based on a sampling of ``Feynman paths'' in Pauli operator space~\cite{aharonov_polynomial-time_2023}, the polynomial scaling of the algorithm features a large exponent (proportional to $1/\varepsilon$, i.e. of order 100 in present-day experiments) that makes the algorithm impractical. 
This leaves open the question of ``practical hardness'' for finite-sized RCS experiments. 
Furthermore, the requirement of anticoncentration in Ref.~\cite{aharonov_polynomial-time_2023} is not met at constant depth $T$ (depth diverging as at least $\log(N)$ is needed~\cite{dalzell_anticoncentrate_2022}), so that even the in-principle hardness of sampling noisy shallow circuits is an open problem.
Our results contribute to sharpen the requirements for such hardness by identifying a phase in the parameter space of depth $T$ and noise strength $\varepsilon$ [Fig.~\ref{fig:idea}(d)] where noisy RCS can be classically simulated via a straightforward MPS algorithm in time\footnote{This scaling corresponds to an MPS with a number $\propto L_x$ of tensors, of constant bond dimension and physical dimension $\propto 2^{cT}$, evolved for $\propto L_y$ steps.} $\sim N \exp(T)$. 
We have located the entanglement phase transitions in circuit architectures based on real-world quantum processors. In square lattices with native iSWAP-like gates, we found that noisy-SEBD allows the efficient sampling of circuits of depth $T = 3$, like SEBD in the noiseless case; but for realistic noise rates of $\varepsilon \lesssim 1\%$ it does not increase the depth threshold (i.e., $T = 4$ remains in the hard phase).
On heavy-hexagon lattices with native CNOT gates, we have instead found that the inclusion of realistic noise rates can increase the depth threshold, as shown in Fig.~\ref{fig:ibm}(c,d). 

Our results add to the growing body of work on noise-induced phase transitions in computational complexity. 
Recent works have studied the simulation of noisy RCS via MPS simulations truncated to constant bond dimension $\chi$~\cite{zhou_what_2020,ayral_density-matrix_2023}, finding that the accumulated truncation error behaves similarly to noise in the quantum experiment---i.e. causes an exponential decay of the linear cross entropy, Eq.~\eqref{eq:XEB}; to beat this classical simulation method, the quantum processor must be below a finite noise threshold. 
We remark that the task considered in those works is different from the one considered here. Namely the goal in Refs.~\cite{zhou_what_2020,ayral_density-matrix_2023} is to simulate the {\it ideal} (noiseless) bitstring distribution better than the noisy quantum experiment, as quantified e.g. by fidelity or linear cross-entropy. Notably this allows for an exponentially-small (in $N$ and $T$) fidelity between simulation and experiment, as long as the former is closer to the ideal result. 
In contrast, our goal is to simulate with high accuracy the {\it noisy} bitstring distribution itself. This is a significant distinction physically as the effect of uncontrolled MPS truncation is {\it a priori} very different from that of local noise, even at the same level of fidelity (e.g., MPS truncation does not obey locality). 

Even more recently, a noise-induced phase transition in RCS has been reported~\cite{morvan_phase_2023,ware_sharp_2023} in deep quantum circuits with noise rates scaled as $\varepsilon \sim 1/N$, i.e. a constant number of errors per layer. The scaling of linear cross entropy [Eq.~\eqref{eq:XEB}] in these models was predicted to sharply change as a function of $\varepsilon N$, from an ``easy phase'' where the system appears to break into finite-sized clusters from the point of view of linear cross-entropy, to a phase where it behaves as a single large cluster. The former phase is easy to spoof classically, while the latter is conjectured to be practically-hard. This transition is also different from the one studied here. For one, it applies to a vanishing noise rate $\varepsilon = O(1/N)$ rather than a finite $\varepsilon = O(1)$. Additionally, and more importantly, it is a phase transition in an observable (albeit a complex one like linear cross-entropy) that reflects an intrinsic property of the system, whereas the transition studied here is a property of a simulation algorithm (noisy-SEBD) that is not intrinsic to the system. 

A distinctive aspect of the problem we study here is that the simulation task requires precise knowledge of the noise model on the quantum processor, which is not fully controlled or programmable, and whose characterization is a separate challenge~\cite{harper_efficient_2020}. For this reason, the sampling task simulated by noisy-SEBD is quite different from the standard (noiseless) one. In other words, {\it given a noise model}, there is a fully well-defined simulation task; however, comparison with real-world noisy devices is more subtle, as the ``true'' noise model for such devices is never completely known. At the same time, this feature of the problem (which is generic of simulation algorithms for noisy systems) opens up an interesting direction for future research, namely using noisy-SEBD as a noise benchmarking tool. One could run noisy-SEBD (or other algorithms for the same task) with a parametrized noise model, compare its output with that of real quantum hardware, and use the comparison to optimize the noise parameters. To this end, it would be useful to generalize our discussion to more complex noise models beyond uncorrelated single-qubit errors. We leave these ideas as directions for future research.

Finally, we emphasize again that noisy-SEBD is only one possible strategy for classical sampling. It remains an interesting open question to identify other simulation algorithms with polynomial cost $O(N^c)$, with $c$ an $\varepsilon$-independent constant (unlike in the Feynman path sampling approach of Ref.~\cite{aharonov_polynomial-time_2023} or the direct MPO approach for 1D circuits of Ref.~\cite{noh_efficient_2020}) below a finite noise threshold $\varepsilon_c$.
The possibility of an intrinsically hard phase at finite depth (before anticoncentration and the results of Ref.~\cite{aharonov_polynomial-time_2023} apply) is also an interesting open question, with noisy-SEBD providing a constraint on this hypothetical phase boundary.
Another interesting direction for future work is the possibility of extending these results to higher dimension or all-to-all connected systems, e.g. trapped ion quantum computers. While MIPTs are known to arise in all these cases, the ensuing area-law for entanglement can only be exploited for efficient MPS simulation in 1D (including in the effective 1D subsystems of shallow 2D circuits used in SEBD). Whether other formulations of the MIPT, e.g. the dynamical purification approach~\cite{gullans_dynamical_2020,gullans_scalable_2020}, can be exploited for efficient simulation in more general systems is an interesting open question. 

{\it Note added.} Upon completion of this manuscript, we became aware of an independent work on related topics appearing in the same arXiv posting~\cite{chen_optimal_2023}. Our works are complementary and our results agree where they overlap. 

\acknowledgments

M. I. thanks Aram Harrow for helpful discussions. We thank Soonwon Choi for bringing Ref.~\cite{kolodrubetz_optimality_2023} to our attention.
Numerical simulations were carried out on computational resources provided by Texas Advanced Computing Center (TACC) and on Stanford Research Computing Center's Sherlock cluster. M. I. was partly supported by the Gordon and Betty Moore Foundation's EPiQS Initiative through Grant GBMF8686.

\appendix

\section{Derivation of the unraveling cost function from statistical mechanical model}\label{ap: statmech}
Here we study the most disentangling unraveling of a single-qubit noise channel $\Phi$ in the context of 1D brickwork random circuits, by mapping the entropy calculation of random circuits to a classical statistical mechanical model~\cite{bao_theory_2020,napp_efficient_2022}. 

\subsection{Quasientropy}
Consider a subsystem $A$ of a one-dimensional qudit chain. The $k$-Renyi entropy for the reduced density matrix $\rho_A$ is defined as
\begin{align}
	S_{k}\(A\)=\frac{1}{1-k}\log\(\frac{Z_{k, A}}{Z_{k,\varnothing}}\),
\end{align}
where
\begin{align}
	Z_{k, \varnothing}=\Tr(\rho)^k,\quad Z_{k, A}=\Tr(\rho_A^k).
\end{align}
The von Neumann entropy is given by the $k\rightarrow1$ limit
\begin{align}
	S_1(A)=-\Tr\(\frac{\rho_A}{\tr\rho}\log\frac{\rho_A}{\tr\rho}\).
\end{align}
For a hybrid random circuit, we are interested in the trajectory-averaged behavior of $k$-Renyi entropy
\begin{align}
	\langle S_k(A)\rangle =\frac{\mathbb{E}_C\[\Tr\rho S_k(A)\]}{\mathbb{E}_C\[\Tr\rho\]}=\frac{1}{1-k}\frac{\mathbb{E}_C\[\Tr\rho \log\frac{Z_{k,A}}{Z_{k, \varnothing}}\]}{\mathbb{E}_C\[\Tr\rho\]},
\end{align}
where $\mathbb{E}_C$ represents the combined average of Haar random circuits $\mathbb{E}_U$ and the average over Kraus operators $\mathbb{E}_M$ (i.e. quantum trajectories).

Replica tricks can be applied to cure the average of the logarithm. However, to get direct mapping to the stat-mech model, one can alternatively consider the $k$-quasientropy $\Tilde{S}_k(A)$~\cite{napp_efficient_2022}, i.e. the $k$-th moment of the entanglement spectrum, weighted by the $k$-th power of the measurement outcome probability,
\begin{align}
	\Tilde{S}_k(A)=&\frac{1}{1-k}\log\(\frac{\mathbb{E}_C\[\tr\(\rho\)^k\frac{Z_{k,A}}{Z_{k, \varnothing}}\]}{\mathbb{E}_C\[\tr\(\rho\)^k\]}\)\nonumber\\=&\frac{1}{1-k}\log\(\frac{\mathbb{E}_C\[Z_{k,A}\]}{\mathbb{E}_C\[Z_{k, \varnothing}\]}\).
\end{align}
Importantly, in the limit $k\rightarrow1$, $\Tilde{S}_k(A)\rightarrow \langle S_1(A)\rangle$, similar to the $k$-Renyi entropy. The $k$-quasientropy has a natural mapping with a classical stat-mech model: the quantities $\mathbb{E}_C[Z_{k,A/\varnothing}]$ can be viewed as partition functions for a (2+0)-dimensional spin model with different boundary conditions.

\subsection{Generalized measurement}
A general measurement procedure can be represented by a set of Kraus operators $\{M_i\}$ satisfying $\sum_i M^\dagger_i M_i=\mathbb{I}$~\cite{nielsen_quantum_2000}. For each quantum trajectory, a specific Kraus operator is chosen with probability $p_i=\tr\(M_i\rho M_i^\dagger\)$ and gives the updated state $\rho'={M_i\rho M^\dagger_i} / {p_i}$. However, in the calculation of the $k$-quasientropy, it is more convenient to re-parametrize the generalized measurement  in terms of operators $\tilde{M}_i$ and classical probabilities $\mu_i$ satisfying
\begin{align}
	\tr\(\Tilde{M}^\dagger_i\Tilde{M}_i\)=q,
	\qquad 
	\mathbb{E}_{\v{M}}M^\dagger M=\sum_i\mu_i\Tilde{M}^\dagger_i \Tilde{M}_i = \mathbb{I},
\end{align}
where $q$ is the local Hilbert space dimension.
It is easy to show $\mu_i$ is given by
\begin{align}
	\mu_i=\frac{1}{q}\tr\(M^\dagger_iM_i\),
\end{align}
and that $\mu_i\geq 0$ and $\sum_i\mu_i=1$ (i.e. $\mu_i$ is a probability distribution).

The advantage of this reparametrization is that it renders the $k$-quasientropy invariant under trivial decomposition of Kraus operators, which is important for optimizing the unraveling. As an example, consider the sets $ \v{M}_0 = \{\sigma_z\}$ and $\v{M}_1 = \{\sigma_z/\sqrt{2}, \sigma_z/\sqrt{2}\}$, which differ by a trivial decomposition of $\sigma_z$ and are hence physically equivalent (both describe the unitary transformation $\rho \mapsto \sigma^z \rho \sigma^z$). 
However, if we define the average over a Kraus set as 
$\mathbb{E}_{\v{M}}[f] = \sum_{M\in \v{M}} f(M \rho M^\dagger)$, then one can verify $\mathbb{E}_{\v{M}_0}[Z_{k,\varnothing}]=1$ but $\mathbb{E}_{\v{M}_1}[Z_{k,\varnothing}] = 2^{1-k}$. 
Using the reparametrized Kraus operators and defining the average as $\mathbb{E}_{\v{M}}[f] = \sum_i \mu_i f(\tilde{M}_i \rho \tilde{M}_i^\dagger)$ instead yields the consistent results $\mathbb{E}_{\v{M}_0}[Z_{k,\varnothing}] = \mathbb{E}_{\v{M}_1}[Z_{k,\varnothing}]=1$. 
It is also worth mentioning that, in the limit of interest ($k\rightarrow1$), these two formalisms are equivalent.

\begin{figure*}[ht]
	\centering
	\includegraphics[width=\textwidth]{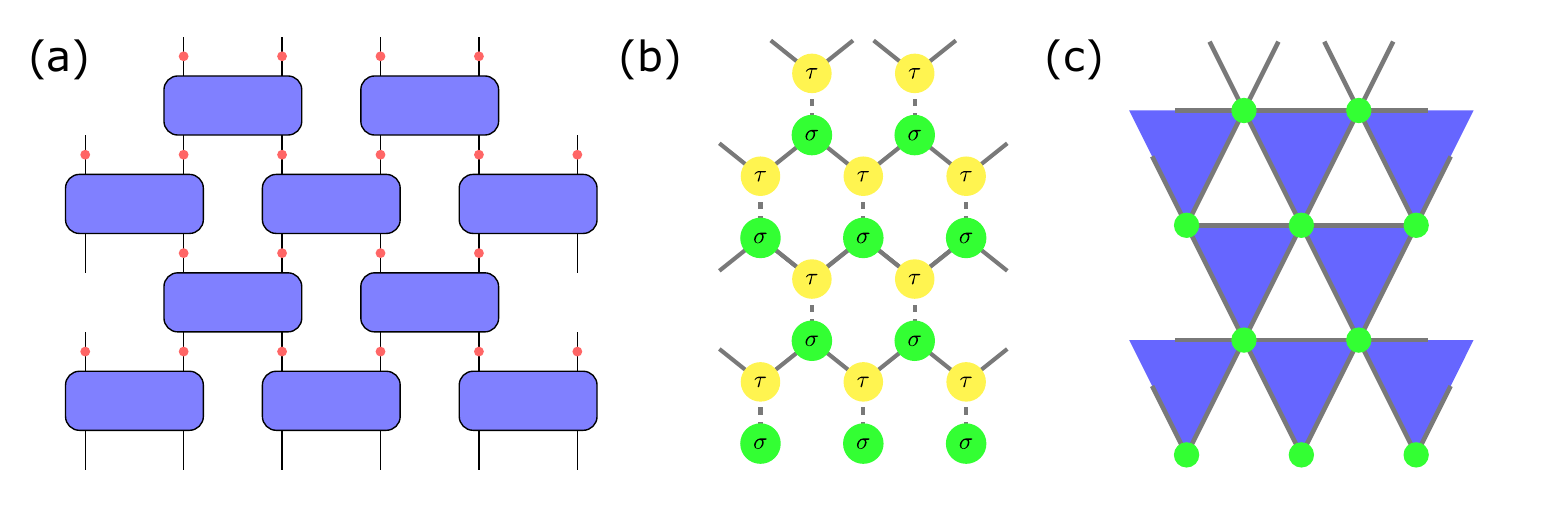}
	\caption{(a) The quantum circuit consists of brick-wall unitaries (blue rectangles) and generalized measurements (red dots). (b) The mapped classical statistical mechanical model on the honeycomb lattice, where dashed links are weighted by the Weingarten function and the solid links are weighted by the generalized measurements. (c) Classical Ising model for $k=2$, after integrating out $\tau$ nodes.}
	\label{fig:statmech}
\end{figure*}

\subsection{Mapping to a classical statistical mechanical model}
The goal of this mapping is to calculate the averaged partition functions $\mathbb{E}_C\[Z_{k,X}\]$ where $X=A$ or $\varnothing$:
\begin{align}
	\mathbb{E}_C\[Z_{k,X}\] = \mathbb{E}_U\mathbb{E}_{\v{M}}
	\[\tr\(\(\cdots MU\rho_0U^\dagger M^\dagger\cdots\)^{\otimes k}\mathcal{S}^{\otimes k}_X\)\],
\end{align}
where $\mathcal{S}_X$ is the operator that implements a cyclical permutation of the $k$ replicas only in the $X$ subsystem. By using the Haar measure calculus~\cite{kostenberger_weingarten_2021}, the average over replicated gates $(U\otimes U^\ast)^{\otimes k}$ can be expanded onto permutations of the replicas, giving a partition function of ``spins'' valued in the permutation group of $k$ elements $S_k$.
Hence the average over single-site measurement operators can be evaluated by appropriately contracting with connecting permutations. 
In this sense, each unitary gate can be viewed as two permutation nodes $\{\sigma, \tau\}\in S_k$, which form a honeycomb lattice as shown in Fig.~\ref{fig:statmech}(b). The total partition function can be evaluated by contracting the nodes with proper weights $w^{(k)}\(\sigma_{\bs{r}},\tau_{\bs{r}'}\)$ on the links:
\begin{align}
	\mathbb{E}_C\[Z_{k,X}\]=\sum_{\{\sigma_{\bs{r}}, \tau_{\bs{r}}\}}\prod_{\langle \bs{r}, \bs{r}'\rangle}w^{(k)}\(\sigma_{\bs{r}},\tau_{\bs{r}'}\),
\end{align}
with distinct couplings on dashed and solid links, as shown in Fig.~\ref{fig:statmech}(b).
The average over Haar random unitary gates gives the coupling for dashed links:
\begin{align}\label{eq:unitary_weight}
	w^{(k)}_u\(\sigma, \tau\)=\mathrm{Wg}_{q^2}\(\tau\sigma^{-1}\),
\end{align}
where $\mathrm{Wg}_{q^2}$ is the Weingarten function~\cite{kostenberger_weingarten_2021}. These couplings are independent of measurements. 
The solid links in Fig.~\ref{fig:statmech}(b), on the other hand, are given by
\begin{align}
	w^{(k)}_m(\sigma, \tau) = \mathbb{E}_{\v{M}} \prod_{c}\tr\(\(M^\dagger M\)^{\lambda_c}\),
\end{align}
where $c$ denotes the number of cycles in permutation $\tau\sigma^{-1} \in S_k$ and $\lambda_c$ is the length of cycle $c$. Using the convention above for averaging over Kraus operators, $\mathbb{E}_{\v{M}}[f] = \sum_i \mu_i f(\tilde{M}_i \rho \tilde{M}_i)$, we obtain
\begin{align} 
	w^{(k)}_m(\sigma, \tau) 
	& = \sum_i\mu_i\prod_c\tr[ ( \Tilde{M}_i^\dagger\Tilde{M}_i)^{\lambda_c}] \nonumber \\
	& =\sum_i\mu_i^{1-k}\prod_c\tr[ ( M_i^\dagger M_i )^{\lambda_c} ].
	\label{eq:measurement_weight}
\end{align}

\subsection{Two replicas: classical Ising model}
Although the combination of Eq.~\eqref{eq:unitary_weight} and Eq.~\eqref{eq:measurement_weight} gives the exact expression for the partition function, the possible negative weights of $w_u(\sigma, \tau)$ impede the direct mapping to a physical system with real interactions at a real temperature. For the case of $k=2$, this sign problem can be circumvented by integrating out all $\tau$ variables, which gives rise to a classical Ising model defined on a triangular lattice as shown in Fig.~\ref{fig:statmech}(c):
\begin{widetext}
	\begin{align}
		\mathbb{E}_C\[Z_{k,X}\]=\sum_{\{\sigma\}}\prod_{\langle\sigma_1, \sigma_2, \sigma_3\rangle}\[\sum_{\tau=\pm1}w_m^{(2)}(\sigma_1, \tau)w_m^{(2)}(\sigma_2, \tau)w_u^{(2)}(\sigma_3, \tau)\]\equiv\sum_{\{\sigma\}}\prod_{\langle\sigma_1, \sigma_2, \sigma_3\rangle}w^{(2)}\(\sigma_1,\sigma_2,\sigma_3\),
	\end{align}
\end{widetext}
where $\langle\sigma_1, \sigma_2, \sigma_3\rangle$ denotes a lower-facing triangle with three neighboring vertices $\sigma_1, \sigma_2, \sigma_3$. For $k=2$, we have
\begin{align}
	w_u^{(2)}\(\sigma, \tau\)=\left\{\begin{aligned}
		&\frac{1}{q^4-1}&\mathrm{if }\ \sigma=\tau,\\
		-&\frac{1}{q^2\(q^4-1\)}&\mathrm{if }\ \sigma\neq\tau,
	\end{aligned}\right.
\end{align}
and denote 
\begin{align}
	w_m^{(2)}\(\sigma, \tau\)=\left\{\begin{aligned}
		&u&\mathrm{if }\ \sigma=\tau,\\
		&v&\mathrm{if }\ \sigma\neq\tau,
	\end{aligned}\right.
\end{align}
where $u$ and $v$ are determined by specific Kraus operators via Eq.~\eqref{eq:measurement_weight}. 
Using the definition of $\mu_i$, one can see that 
\begin{align}
	u &= \sum_i\mu_i^{-1}\tr\(M_i^\dagger M_i\)^2=q^2,\label{eq:u}\\
	v&=\sum_i\mu_i^{-1}\tr\(\(M_i^\dagger M_i\)^2\)=q\sum_i\frac{\tr\(M_i^\dagger M_iM_i^\dagger M_i\)}{\tr\(M_i^\dagger M_i\)}.\label{eq:v}
\end{align}

With these, one can express the three-body interaction $w^{(2)}\(\sigma_1, \sigma_2, \sigma_3\)$ explicitly. Due to the permutation symmetry $\sigma\rightarrow \bar{\sigma}$ (where $\bar{\sigma}$ denotes the other permutation with $S_2\equiv \mathbb{Z}_2$) and spatial reflection symmetry $\sigma_1\leftrightarrow\sigma_2$, one only needs to specify
\begin{align}
	w^{(2)}\(\sigma, \sigma, \sigma\)=&\frac{u^2}{q^4-1}-\frac{v^2}{q^2\(q^4-1\)},\\ w^{(2)}\(\sigma, \sigma, \bar{\sigma} \)=&\frac{v^2}{q^4-1}-\frac{u^2}{q^2\(q^4-1\)},\\ w^{(2)}\(\sigma, \bar{\sigma}, \sigma\)=&\frac{uv}{q^2\(q^2+1\)}.
\end{align}
Furthermore one may reexpress the three-body interaction as the product of three two-body interaction $w^{(2)}\(\sigma_1, \sigma_2, \sigma_3\)=Ce^{-J_h\sigma_1\sigma_2-J_d\sigma_1\sigma_3-J_d\sigma_2\sigma_3}$, where $J_d$ and $J_h$ are the two-body interaction strength for diagonally and horizontally neighboring sites respectively, whose expressions are
\begin{align}
	J_d &= \frac{1}{4}\log\(\frac{q^2x^2-1}{q^2-x^2}\),\\
	J_h&=\frac{1}{4}\log\(\frac{x^2\(q^2-1\)^2}{\(q^2x^2-1\)\(q^2-x^2\)}\),
\end{align}
where $x\equiv v/u$ is a dimensionless parameter given by
\begin{equation}
	x = \frac{1}{q}\sum_i\frac{\tr\(M_i^\dagger M_iM_i^\dagger M_i\)}{\tr\(M_i^\dagger M_i\)}. \label{eq:app_x_def}
\end{equation}
Since $\tr\(M^\dagger M\)^2\geq\tr\(\(M^\dagger M\)^2\)$ for all Kraus operators, one has $u\geq v$, hence $J_d\leq0$ and $J_h\geq0$ for $q\geq2$. 

The statistical mechanics of this classical Ising model can be solved exactly~\cite{eggarter_triangular_1975,tanaka_triangular_1978} and the critical point $x_c$, separating paramagnet and ferromagnet phases, is determined by the relation $2e^{2J_h}=e^{-2J_d}-e^{2J_d}$. Solving for $x_c$ gives 
\begin{align}
	\frac{1}{x_c}=\frac{q^2-1}{q^2+1}+\frac{\sqrt{2q^4+2}}{q^2+1}.
\end{align}
This phase transition between an ordered and disordered phase in the statisitical mechanical model then maps onto to the area-law/volume-law phase transition of the $2$-quasientropy in the hybrid random circuit.
As we see, within this two-replica analysis the transition is controlled exclusively by the parameter $x$, Eq.~\eqref{eq:app_x_def}, with an ordered phase (volume-law) for $x < x_c$ and disordered phase for $x > x_c$ (area-law). Therefore, to obtain the optimal unraveling for a given quantum channel, we aim to maximize the function $x(\v{M})$. This justifies the use of $x(\v{M})$ as a cost function for the unraveling of $\Phi$ in the main text. 

\begin{figure*}
	\centering
	\includegraphics[width=\textwidth]{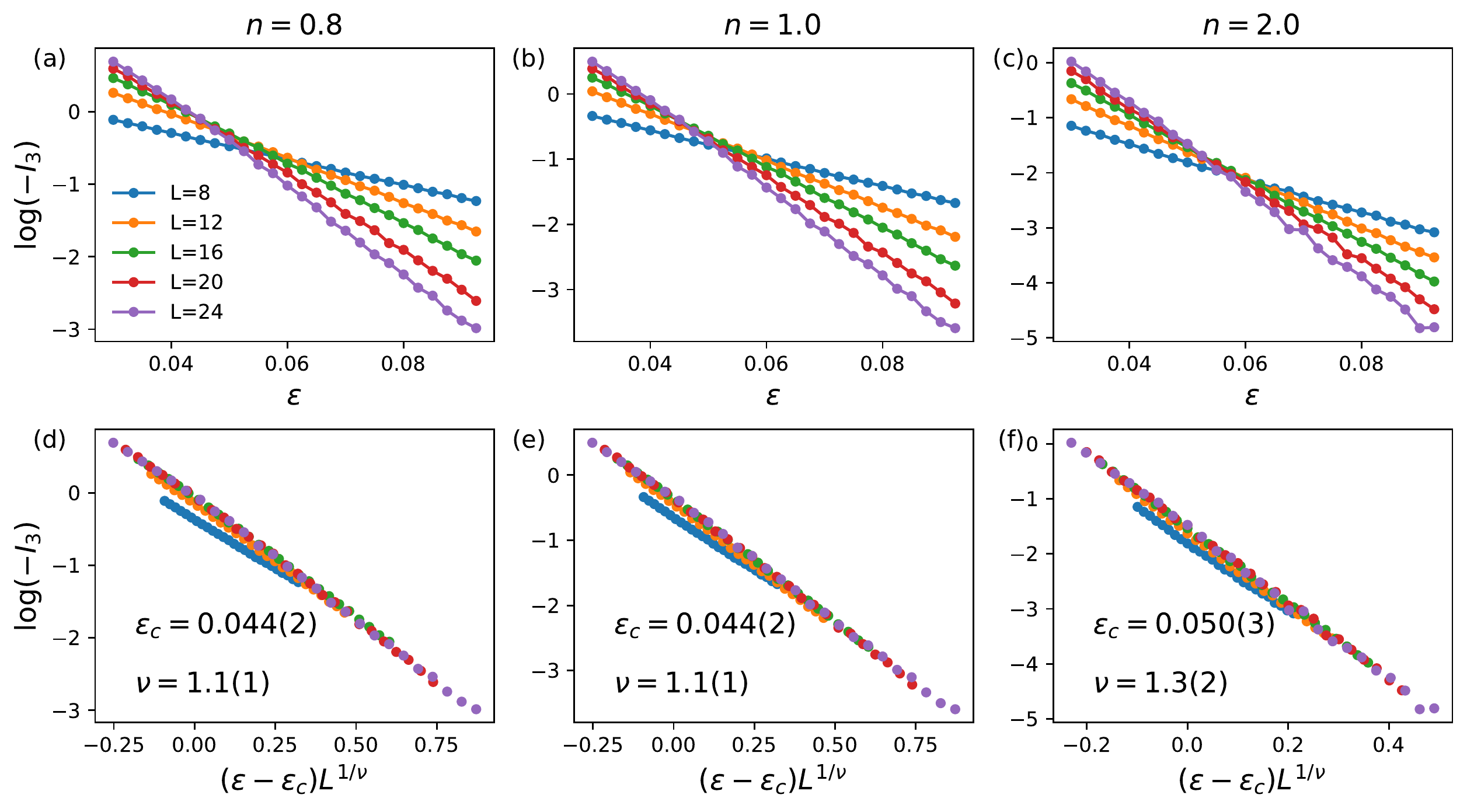}
	\caption{(a-c) Tripartite mutual information $I_3$ (shown as $\log(-I_3)$) in 1D random circuits with dephasing noise of strength $\varepsilon$ unraveled into optimal weak measurements, for Renyi indices (a) $n=0.8$, (b) $n=1.0$, and (c) $n=2.0$.
	(d-f) Data collapse of $\log(-I_3)$ vs $(\varepsilon - \varepsilon_c) L^{1/\nu}$, with fit parameters indicated. The data are averaged over $5\cdot10^2-5\cdot10^3$ realizations}
	\label{fig:I3_renyi}
\end{figure*}

\section{Maximization of the target function for unital channels \label{ap:inequality}}

To compute Eq.~\eqref{eq:targetfunction} for the parametrizaion of Kraus operators in Eq.~\eqref{eq:Krausoperatoransatz}, we first evaluate $M_i^\dagger M_i$. Omitting the subscript $i$ for ease of notation, we have
\begin{equation}
	M^\dagger M = (a^2 + b^2) \mathbb{I} + 2b(a \v{u}_R + b \v{u}_I \wedge \v{u}_R) \cdot \v{\sigma},
\end{equation}
with $\tilde{\v u} = \v{u}_R + i \v{u}_I$. 
(We used the identity $\sigma^\alpha \sigma^\beta = \sigma^0 \delta_{\alpha,\beta} + i\varepsilon_{\alpha\beta\gamma} \sigma^\gamma$.)

Letting $\v{v} \equiv a \v{u}_R + b \v{u}_I \wedge \v{u}_R$, the operator $M^\dagger M$ has eigenvalues $\lambda_\pm = (a^2+b^2) \pm 2b \|\v{v}\|$. It follows that the ratio in the definition of $x(\v{M})$, Eq.~\eqref{eq:targetfunction}, reads
\begin{equation}
	\frac{\Tr[(M^\dagger_i M_i )^2]} {2\Tr(M^\dagger_i M_i )}
	=  \frac{a_i^2 + b_i^2}{2} + \frac{2b_i^2 \| \v{v}_i \|^2}{a_i^2 + b_i^2}.
\end{equation}
Focusing on the second term, we have $\| \v{v} \|^2 = a^2 \cos^2(\theta) + b^2 \cos^2(\theta) \sin^2(\theta) \sin^2(\chi)$, where $\| \v{u}_R\| \equiv \cos(\theta)$ (note $\tilde{\v u}$ is a unit vector, so $\|\v{u}_R\|^2 + \|\v{u}_I\|^2 = 1$) and $\chi$ is the angle between $\v{u}_R$ and $\v{u}_I$. 
Since we aim to maximize $x(\v{M})$, we can take $\chi = \pi/2$ and then maximize over $\theta$ the function $f(\theta) = a^2 \cos^2(\theta) + b^2 \cos^2(\theta) \sin^2(\theta)$.
The maximum depends on the ratio of $a$ and $b$: 
\begin{equation}
	\max_\theta f(\theta) = 
	\left\{ 
	\begin{aligned}
		& a^2 & \text{if } a > b, \\
		& \frac{(a^2+b^2)^2}{4b^2} & \text{if } a \leq b. 
	\end{aligned}
	\right. \label{eq:theta_maximum}
\end{equation}
It follows that 
\begin{equation}
	x(\v{M}) = 1 - \frac{1}{2} \sum_{i:\, a_i > b_i} \frac{(a_i^2 - b_i^2)^2}{a_i^2 + b_i^2},
\end{equation}
where $a_i$ and $b_i$ are subject to the constraints $\sum_i a_i^2 = p_0$ and $\sum_i b_i^2 = 1 - p_0$, Eq.~\eqref{eq:abconstrain}. 
For strong noise, $p_0 \leq 1/2$, it is possible to choose $a_i \leq b_i$ for all $i$. This gets rid of the negative term in the sum and gives $x(\v{M}) = 1$, i.e. complete purification, which is optimal. 
However for weak noise ($p_0 > 1/2$) we have $\sum_ia_i^2>\sum_ib_i^2$, so it is not possible to avoid the sum over $i$ in the above expression. 

We first handle the case in which $a_i > b_i$ for all $i$. 
Introducing $c_i = a_i^2 + b_i^2$ and $d_i = a_i^2 - b_i^2$, we have 
\begin{equation}
	x(\v{M}) 
	= 1 - \sum_i \frac{d_i^2}{2c_i} 
	\leq 1 - \frac{(\sum_i d_i)^2}{2\sum_i c_i},
\end{equation}
where we used Sedrakyan's inequality (i.e. Cauchy-Schwartz applied to the vectors $\{\sqrt{c_i} \}$ and $\{d_i/\sqrt{c_i}\}$). 
The constraints in Eq.~\eqref{eq:abconstrain} impose $\sum_i c_i = 1$ and $\sum_i d_i = 2p_0 - 1$, therefore 
\begin{equation}
	x(\v{M}) \leq 1 - \frac{(2p_0 - 1)^2}{2}.
\end{equation}
This is saturated by setting $a_i = \sqrt{p_0/n}$ and $b_i = \sqrt{(1-p_0)/n}$ for all $i$.

Next, we consider the case in which $a_i \leq b_i$ for some $i$. We use primed sums to denote sums over $i$ that are restricted only to those indices: $\sum_{i:\ a_i \leq b_i} = \sum_i'$. 
Let us define $\gamma \equiv \sum_i' c_i$ (we have $0\leq \gamma \leq 1$, where $\gamma = 0$ recovers the previous case); then, by the same reasoning as above, we have 
\begin{equation}
	x(\v{M}) \leq
	1-\frac{(2p_0 - 1 - \sum_i' d_i)^2}{2(1-\gamma)}.
\end{equation}
Now $-\sum_i' d_i = \sum_i' b_i^2 - a_i^2$ is non-negative by definition, so 
\begin{equation}
	x(\v{M}) \leq 1 - \frac{(2p_0-1)^2}{2(1-\gamma)},
\end{equation}
where the right-hand side is maximized for $\gamma = 0$. 
Thus the symmetric solution $a_i = \sqrt{p_0/n}$ and $b_i = \sqrt{(1-p_0)/n}$ for all $i$ is optimal in general. 

Finally, note that when $a_i > b_i$, the maximum in Eq.~\eqref{eq:theta_maximum} is achieved at $\theta = 0$, i.e. $\|\v{u}_I\| = 0$; thus in the solution above, the unit vectors are real: $\tilde{\v u} = \v{u}_R$.

\section{Measurement-induced phase transition for the optimal weak measurement in 1D random circuits}\label{ap:1Doptimal}
Here we analyze  measurement-induced phase transition corresponding to weak measurements from optimal unraveling in Eq.~(\ref{eq:weakmeasurementsdephasing})
against stochastic projective measurements in Eq.~(\ref{eq:dephasing_to_projective}).

We consider 1D Haar random brick-wall circuits with periodic boundary condition and optimal weak measurements applied between layers of unitaries. We use the tripartite mutual information\footnote{For $n\neq 1$ this quantity is not a proper mutual information.}
\begin{align}
	I_{3,n} =& S_{n}(A) + S_{n}(B) + S_{n}(C) - S_n(A\cup B)\nonumber\\ &- S_n(A\cup C) - S_n(B\cup C)+ S_n(A\cup B\cup C) ,
\end{align}
where $S_n(X)$ is the Renyi entropy and $A, B, C$ are contiguous subsystems of size $L/4$ (i.e. the system is divided into quarters and three such subsystems are used). $I_{3, n}$ was argued to be system-size-independent constant at criticality and thus can be used to accurately locate the critical point~\cite{zabalo_critical_2020}.

In Fig.~\ref{fig:I3_renyi} we show the $I_{3, n}$ for $n=0.8,\ 1.0,\ 2.0$ at late time $t=4L$, varying the system size $L$ from $8$ to $24$.  Since the crossing points drift to lower $\varepsilon$ as system size increases, we only consider the data collapse for $L=16, 20, 24$. The obtained critical point $\varepsilon_{c, {\rm weak}}\approx0.044$ is smaller than $\varepsilon_{c, {\rm projective}}\approx0.084$~\cite{zabalo_critical_2020} (Note that in our convention $\varepsilon$ is interpreted as noise rate which is half of measurement rate in the context of the measurement-induced phase transition: $p  = 2\varepsilon$). This is consistent with our argument that for a given noise rate, optimal weak measurements are more disentangling than projective measurements. 

\begin{figure*}
	\centering
	\includegraphics[width=\textwidth]{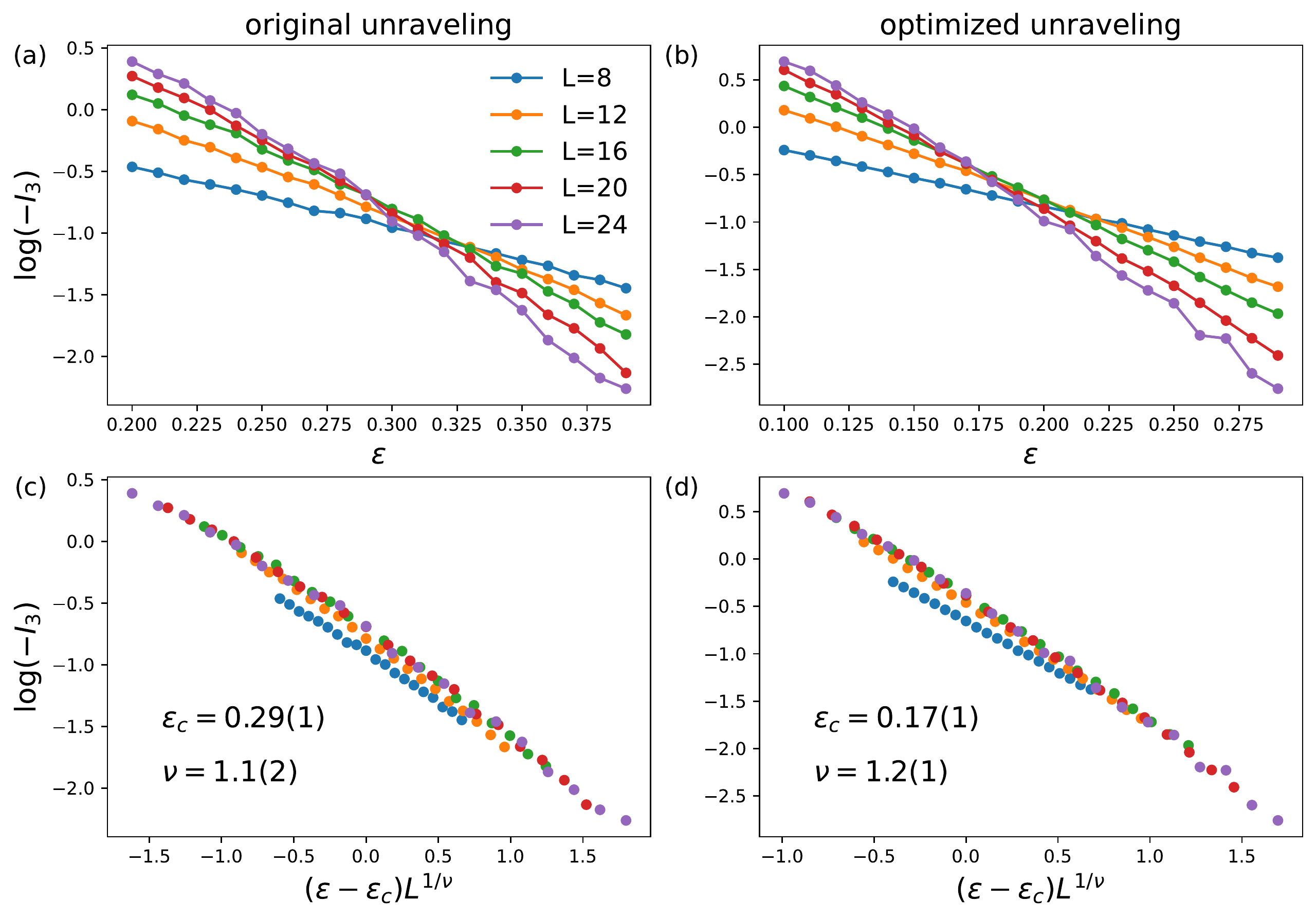}
	\caption{Tripartite mutual information $I_3$ (shown as $\log(-I_3)$) for $n=1.0$ in 1D random circuits with damping of strength $\varepsilon$ for (a) original unraveling and (b) optimized unraveling.
		(c) and (d) Corresponding data collapse of $\log(-I_3)$ vs $(\varepsilon - \varepsilon_c) L^{1/\nu}$, with fit parameters indicated. The data are averaged over $2\cdot10^2-5\cdot10^3$ realizations}
	\label{fig:I3_damp}
\end{figure*}

\section{Optimization for the amplitude damping channel}\label{ap:1Ddamp}
In Sec. \ref{sec:unital}, we give an analytical solution to the optimization for unital qubit channels. Non-unital channels, on the other hand, are generally hard to solve exactly due to the absence of spherical symmetry, thus one usually seeks for a numerical solution using optimization algorithms. However, for a common non-unital channel, the amplitude damping channel with Kraus operators
\begin{align}\label{eq:dampingchannel}
	 \left\{ \begin{aligned}
	 	M_0=&\begin{pmatrix}
	 		1&0\\
	 		0&\sqrt{1-\varepsilon}
	 	\end{pmatrix},\\
	 	M_1 =& \begin{pmatrix}
	 		0&\sqrt{\varepsilon}\\
	 		0&0
	 	\end{pmatrix},
	 \end{aligned}\right.
\end{align}
 the optimization problem can highly simplified since its Kraus rank is only 2. Here we consider the optimization over generic $2\times2$ unitary matrices taking the form
 \begin{align}\label{eq:twounitary}
 	U=e^{i \varphi / 2}\begin{pmatrix}
 		e^{i \psi} & 0 \\
 		0 & e^{-i \psi}
 	\end{pmatrix}\begin{pmatrix}
 		\cos \theta & \sin \theta \\
 		-\sin \theta & \cos \theta
 	\end{pmatrix}\begin{pmatrix}
 		e^{i \Delta} & 0 \\
 		0 & e^{-i \Delta}
 	\end{pmatrix}.
 \end{align}

Substituting Eq.~\eqref{eq:dampingchannel}, \eqref{eq:twounitary} into Eq.~\eqref{eq:targetfunction}, one can obtain
\begin{align}
	x = \frac{1}{2+\frac{4\varepsilon^2}{1+3\varepsilon-(1-\varepsilon)\cos4\theta}},
\end{align}
reaching maximum when $\theta=\pi/4$, which gives the optimized unraveling
\begin{align}\label{eq:dampoptimized}
	\left\{ \begin{aligned}
		{M}_0=&\begin{pmatrix}
			1&\sqrt{\varepsilon}\\
			0&\sqrt{1-\varepsilon}
		\end{pmatrix}/\sqrt{2},\\
		{M}_1 =& \begin{pmatrix}
			-1&\sqrt{\varepsilon}\\
			0&-\sqrt{1-\varepsilon}
		\end{pmatrix}/\sqrt{2}.
	\end{aligned}\right.
\end{align}

We also numerically verify the effect of disentangling from using the optimized unraveling in the context of measurement-induced phase transition in 1D random circuits. In Fig.~\eqref{fig:I3_damp} we show the tripartite mutual information $I_{3}$ at late time $t=4L$ in the 1D Haar random circuits subject to the amplitude damping channel. The critical point for optimized unraveling Eq.~\eqref{eq:dampoptimized} $\varepsilon_{c}\approx0.17$ is much smaller than that for original unraveling Eq.~\eqref{eq:dampingchannel} $\varepsilon_{c}\approx0.29$, indicating the disentangling effect from unraveling optimization also works for non-unital channels.

\begin{figure}
	\centering
	\includegraphics[width=0.5\textwidth]{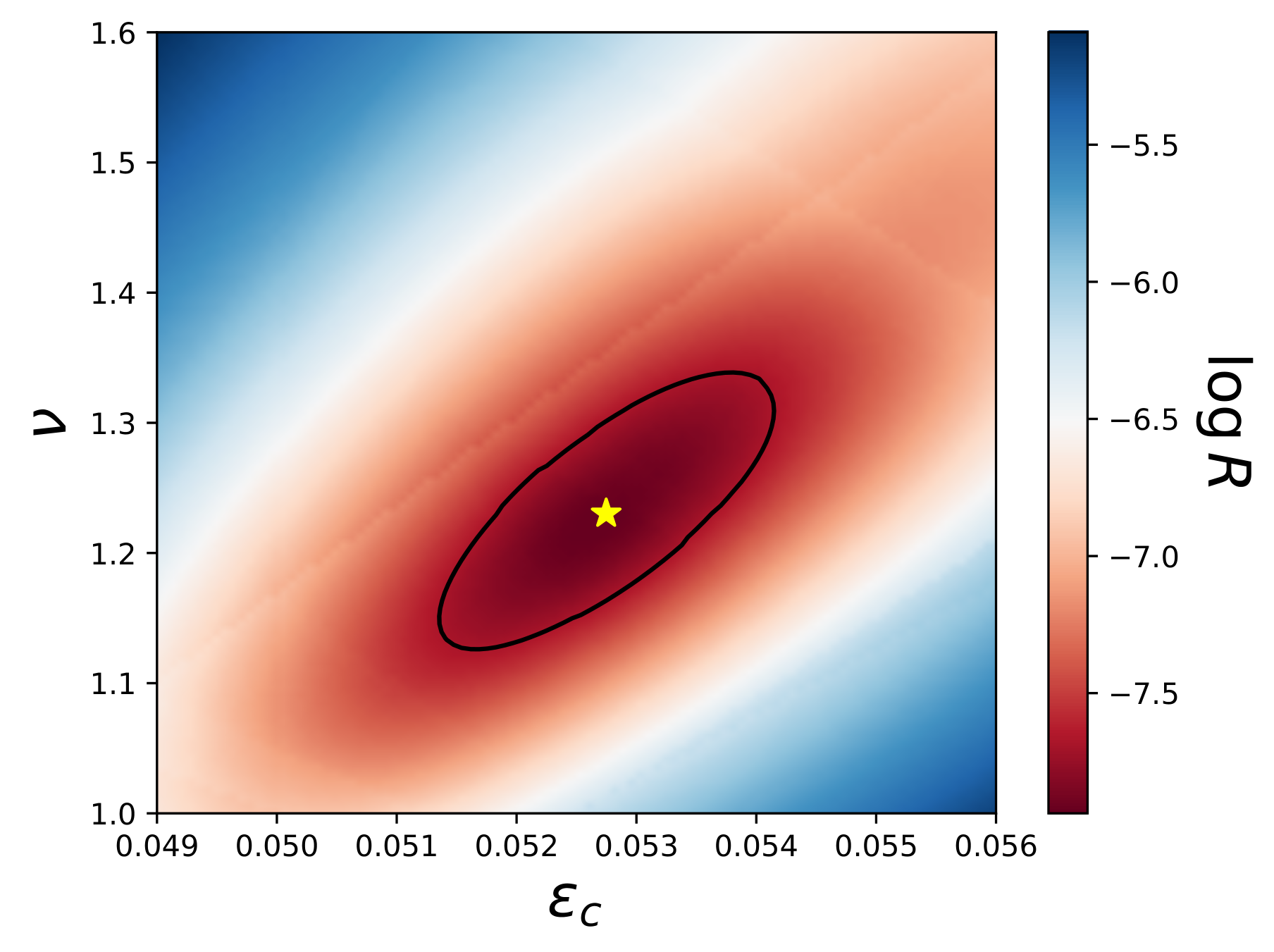}
	\caption{Color map of data collapse object function $R$, Eq.~\eqref{eq:objectivefunction}. The black contour encloses a region satisfying $R\leq1.3\cdot R_{\rm  min}$, which gives an estimation of the error.}
	\label{fig:datacollapse}
\end{figure}

\section{Data Collapse}\label{ap:collapse}
In Sec.~\ref{sec:numerics}
we determine the critical point $\varepsilon_c$ and correlation length exponent $\nu$ by finding a data collapse satisfying the scaling ansatz
\begin{align}
	\tau(\varepsilon, L_x) =L_x F\[\(\varepsilon-\varepsilon_c\)L^{1/\nu}_x\].
\end{align}
To quantify the collapse we consider the similar objective function as in Ref.~\cite{zabalo_critical_2020}. We first sort the data by $x_i=\(\varepsilon_i-\varepsilon_c\)L_{x, i}^{1/\nu}$ where $i=\{1,\cdots, n\}$ labels sorted data points and define $y_i=\tau(x_i)/L_{x,i}$. Then the objective function $R(\varepsilon_c, \nu)$ is defined as
\begin{align}\label{eq:objectivefunction}
	R(\varepsilon_c, \nu) = \frac{1}{n-2}\sum_{i=2}^{n-1}\(y_i-\bar{y}_i\)^2,
\end{align}
where
\begin{align}
	\bar{y}_i=\frac{(x_{i+1}-x_i)y_{i-1}-(x_{i-1}-x_i)y_{i+1}}{x_{i+1}-x_{i-1}},
\end{align}
is the estimation of $y_i$ given by linear interpolating $(x_{i-1}, y_{i-1})$ and $(x_{i+1}, y_{i+1})$.  Then the aim is to minimize Eq.~\ref{eq:objectivefunction} over $\(\varepsilon_c, \nu\)$. As an example, a color plot of $R$ is shown in Fig.~\ref{fig:datacollapse} for data set of $T=5$ (ABCDB) shown in Fig.~\ref{fig:PT} (c). We estimate the error by consider a region enclosed by $R=1.3\cdot R_{\rm min}$, which is shown as a black contour in Fig.~\ref{fig:datacollapse}.

\section{Phase boundary at large $T$ \label{ap:largeT_limit}}

Here we present the results of stabilizer simulations of Clifford circuits to study the entanglement phase transition in circuits of large depth $T$. The goal is to qualitatively study the phase boundary in Fig.~\ref{fig:idea}(d) at values of $T$ that are beyond what can easily be studied via exact simulation of generic circuits (e.g. $T = 4,5$ in Fig.~\ref{fig:PT}).
Due to the restriction of stabilizer simulation, we cannot unravel depolarizing noise into weak measurements. For this reason we use stochastic projective measurements, bearing in mind that this will give a {\it larger} numerical value of the threshold noise strength $\varepsilon_c$ (cf Appendix~\ref{ap:1Doptimal}). We expect the qualitative behavior of $\varepsilon_c(T)$ to be similar across weak and projective unravelings.

\begin{figure}
\centering
\includegraphics[width=\columnwidth]{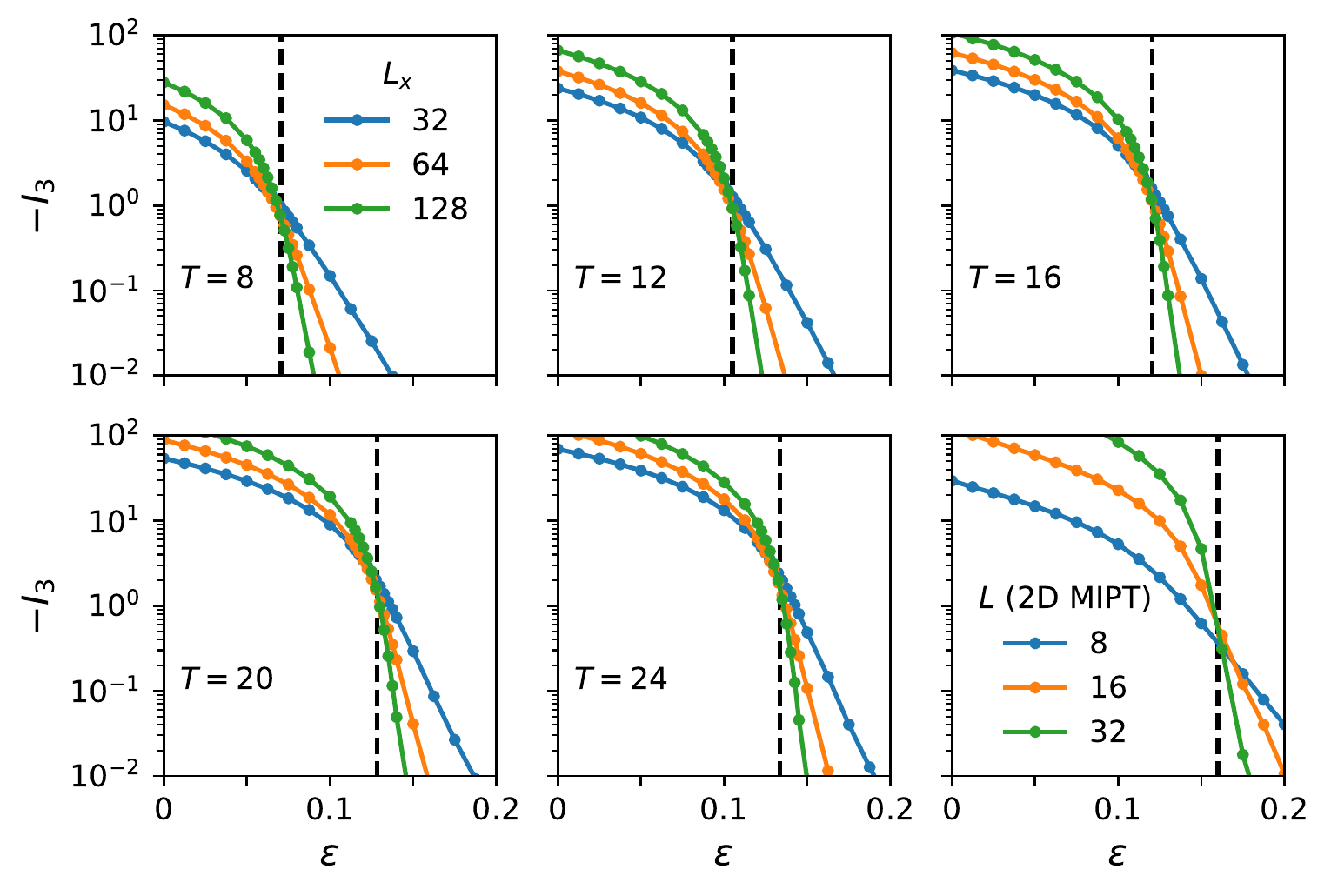}
\caption{Entanglement phase transition in Clifford circuits with dephasing noise of strength $\varepsilon$. First 5 panels refer to quasi-1D subsystems that mimic the noisy-SEBD simulation of 2D circuits of depth $T = 8, 12, 16, 20, 24$ ($T = 4$, not shown, is found to be in the area-law phase for all $\varepsilon$). 
Last panel shows data for the conventional MIPT in 2D square lattices of size $L\times L$, $L = 8, 16, 32$, with circuits of depth $O(L)$. Vertical dashed lines indicate estimates of the critical point.
Data obtained by averaging between $4\times 10^2$ and $2\times 10^4$ realizations of the random circuits depending on system size.
\label{fig:dim_crossover}
}
\end{figure}

\begin{figure}
\centering
\includegraphics[width=\columnwidth]{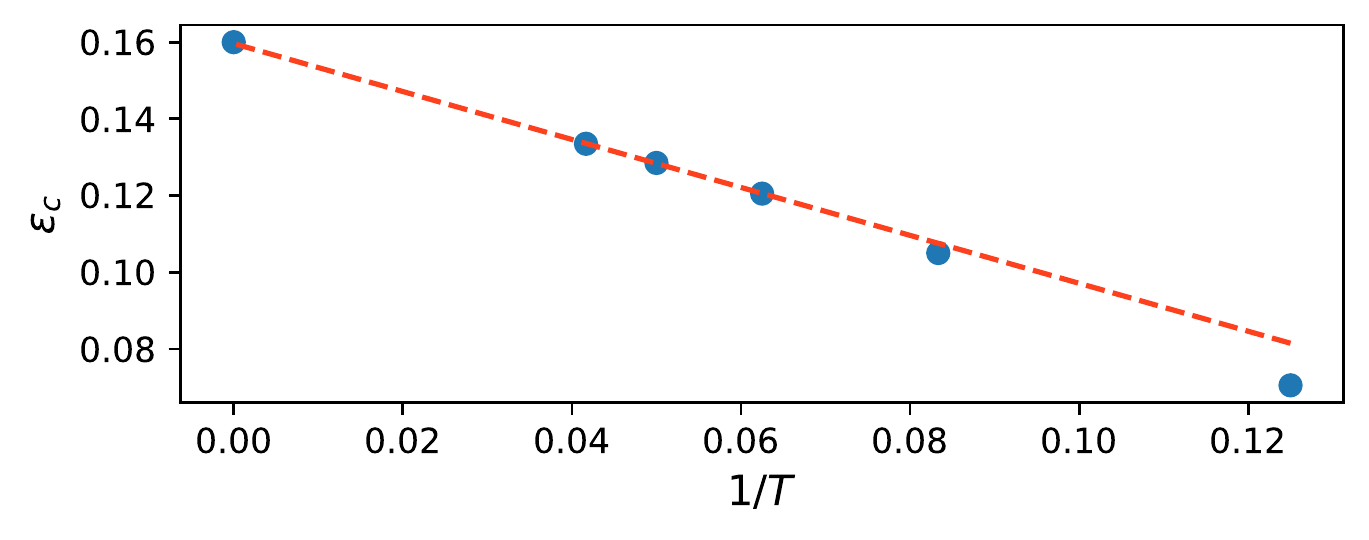}
\caption{Noise threshold for the entanglement phase transition as a function of circuit depth $T$ (data from Fig.~\ref{fig:dim_crossover}). Dashed line indicates a linear fit of datapoints $T = 16,20,24$ to $a + b/T$; the extrapolation to $T = \infty$ is in good agreement with the 2D MIPT datapoint.
\label{fig:epsc_vs_T}
}
\end{figure}

We consider Clifford circuits that mimic the models analyzed in Sec.~\ref{sec:numerics}. To approximate the iSWAP-like gates, we sample two-qubit Clifford gates that are iSWAP with $90\%$ probability and SWAP otherwise (corresponding to a uniform sampling of the {\it dual-unitary} half of the two-qubit Clifford group~\cite{zabalo_operator_2022}). 
These gates are sandwiched between random single-qubit Clifford rotations. A projective measurement of $Z$ is applied with probability $p = 2\varepsilon$ on each qubit after each gate.
Specifically, we simulate the space-wise evolution of the circuit as in e.g. Fig.~\ref{fig:grid}(c). This involves a quasi-1D subsystem which is a strip of length $L_x$ and width $1+T/4$. 
As a diagnostic of the phase transition we use the tripartite mutual information $I_3(A:B:C)$ between contiguous regions that make up strips of length $L_x/4$. 
Results in Fig.~\ref{fig:dim_crossover} show a transition at critical noise rate $\varepsilon_c(T)$ that increases with $T$, as expected. We have, e.g., $\varepsilon_c(T = 8) \simeq 0.070$ which increases to $\varepsilon_c(T = 24) \simeq 0.133$.

We additionally simulate the conventional MIPT in square 2D circuits, whose dimensions $L_x = L_y = L$ are jointly increased. The tripartite mutual information $I_3$, for a partition of the square into 4 rectangles of size $L/4 \times L$, also shows a transition. We observe $\varepsilon_{c,2D} \simeq 0.16$, consistent with the value $p_{c,2D} = 0.3116(1)$ reported in Ref.~\cite{sierant_measurement-induced_2022} for this circuit architecture\footnote{See Appendix A.1 therein. While the circuit architecture is the same, the gate set is slightly different---random Clifford gates, not restricted to SWAP and iSWAP.} (recall $p = 2\varepsilon$).

Finally, in Fig.~\ref{fig:epsc_vs_T} we compare the observed critical points for shallow circuits of depth $T$, $\varepsilon_c(T)$, with the 2D MIPT $\varepsilon_{c,2D}$. We see good agreement with the conjectured form $\varepsilon_c(T) \simeq \varepsilon_{c,2D} + O(1/T)$, sketched in Fig.~\ref{fig:idea}(d).

\begin{figure*}
	\centering
	\includegraphics[width=\textwidth]{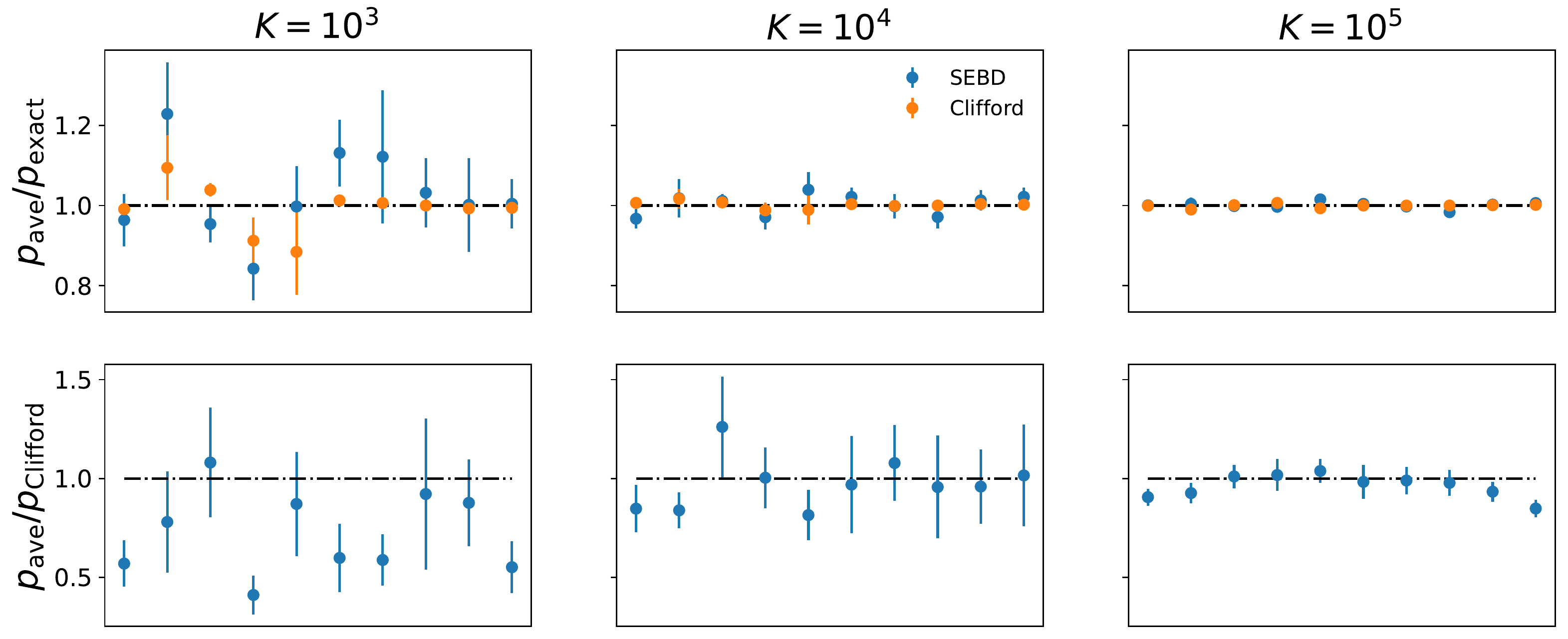}
	\caption{Estimated probability of $10$ randomly generated output bitstrings of noisy Clifford circuits of depth $T=4$, with $L_x=5, L_y=5$ (top row) and $L_x=9, L_y=9$ (bottom row). Probabilities are estimated by averaging over $K$ trajectories, with $K = 10^3$-$10^5$ indicated on top.
		In the top row, the averaged probabilities $p_{\rm ave}$ (from noisy-SEBD and stabilizer simulations) are normalized by the exact value obtained from MPO simulation. 
		In the bottom row, the averaged probabilities $p_{\rm ave}$ from noisy-SEBD are normalized by the average of $10^6$ trajectories of stabilizer simulations. 
		The ratios converge towards 1 (horizontal lines) with increasing $K$.}
	\label{fig:clifford}
\end{figure*}
\section{Benchmarks}\label{ap:benchmark}
Here we benchmark our noisy-SEBD algorithm against stabilizer simulations of Clifford random circuits and MPO simulation with controlled errors. Since the depolarizing channel with general $\varepsilon$ is not a Clifford operation, we consider the probabilistic trace setup~\cite{li_entanglement_2023} i.e. replacing the depolarizing channel by a probabilistic mixture of an identity operation with probability $1-\frac{4}{3}\varepsilon$ and a trace channel (or erasure) $\rho \mapsto (\mathbb{I}/2)_i \otimes \Tr_i\rho$ with probability $\frac{4}{3}\varepsilon$. With this replacement, each instance/trajectory in this random ensemble is classically simulatable~\cite{aaronson_improved_2004}, and on average reproduces the effect of the depolarizing noise with strength $\varepsilon$.

We first consider the noisy sampling problem in the architecture given in Sec.~\ref{sec:numerics} with $\varepsilon=0.02$, $L_x=5$, $L_y=5$ and $T=4$.
In this case the system size is small enough that direct MPO simulation of the density matrix can be implemented with small error and can thus serve as a benchmark for noisy-SEBD. 
In the upper row of Fig.~\ref{fig:clifford}, we show the averaged final probability $P_{\mathcal N}(\v{z})$ of $10$ randomly chosen output bitstrings $\v z$ from noisy Clifford circuits, varying the number of sampled trajectories $K$ from $10^3$ to $10^5$ (note we unravel depolarizing noise into probabilistic erasure for the stabilizer simulation, and into weak measurements for noisy-SEBD; we use the same $K$ for both methods). The probabilities $P_{\mathcal N}(\v{z})$ are scaled by the exact reference value obtained from MPO simulation, whose truncation error is kept below $10^{-10}$. As the trajectory number increases, one can see both SEBD and Clifford simulation show a good convergence to the exact result. 

We then consider the same circuit architecture but with $L_x=9$, $L_y=9$, where exact MPO computation would require large computational effort. Therefore, in this case we directly compare SEBD results against the Clifford results (with $10^6$ trajectories for the latter). Results are shown in the bottom row of Fig.~\ref{fig:clifford}, again displaying good agreement.

\section{Converting gate fidelity to noise rate \label{ap:fidelity}}

The convention for noise strength $\varepsilon$ used in this work is in terms of single-qubit channels as in Eq.~\eqref{eq:depolarize}. The usual figure of merit for two-qubit gates is the average fidelity $f$. To convert between the two, we note that
\begin{align}
f & = \int {\rm d}\psi\, \bra{\psi} \Phi^{\otimes 2}[ \ketbra{\psi}] \ket{\psi} \nonumber \\
& = (1-\varepsilon)^2 + [1-(1-\varepsilon)^2] \int {\rm d}\psi\, \bra{\psi} P \ket{\psi}^2,
\end{align}
where the integral is over the Haar measure on the two-qubit Hilbert space, $P$ is any traceless Pauli operator, and the formula applies equally to dephasing and depolarizing noise (in fact to any unital noise channel upon setting $1-\varepsilon \mapsto p_0$, cf Eq.~\eqref{eq:unitalchannel}).
The Haar integral yields $\Tr(P^2)/20 = 1/5$, so
\begin{equation}
f = (1-\varepsilon)^2 + \frac{\varepsilon(2-\varepsilon)}{5} \simeq 1 - \frac{8}{5} \varepsilon,
\end{equation}
at small $\varepsilon$, thus $\varepsilon \simeq (5/8)(1-f)$. 
In particular this means that a gate fidelity of $f = 96\%$ translates to $\varepsilon \simeq 0.025$, which is used in Sec.~\ref{sec:ibm}.

\bibliography{nSEBD}

\end{document}